\newcommand{\Ntotal}  {3068}
\newcommand{\intLdt}  {183~}
\newcommand{\intLdtfull}  {183.05}
\newcommand{\dLtot}   {0.40}
\newcommand{\rroots}  {188.64}  
\newcommand{\droots}  {0.04} 
\newcommand{\sysdkg} {\mbox{$-0.03^{+0.26}_{-0.19}$}} 
\newcommand{\syslam} {\mbox{$-0.118^{+0.066}_{-0.064}$}}
\newcommand{\sysdgz} {\mbox{$-0.018^{+0.067}_{-0.064}$}}
\newcommand{\ioddkg} {\mbox{$-0.00^{+0.23}_{-0.18}$}} 
\newcommand{\iodlam} {\mbox{$-0.113^{+0.059}_{-0.057}$}}
\newcommand{\ioddgz} {\mbox{$-0.008^{+0.059}_{-0.058}$}}
\newcommand{\resdkg} {\mbox{$-0.03^{+0.20}_{-0.16}$}} 
\newcommand{\reskg} {\mbox{$0.97^{+0.20}_{-0.16}$}} 
\newcommand{\reslam} {\mbox{$-0.110^{+0.058}_{-0.055}$}}
\newcommand{\resdgz} {\mbox{$-0.009^{+0.060}_{-0.057}$}}
\newcommand{\resgz} {\mbox{$0.991^{+0.060}_{-0.057}$}}
\newcommand{\dkgint} {\mbox{$[-0.32,\;\;0.45]$}} 
\newcommand{\lamint} {\mbox{$[-0.22,\;\;0.01]$}} 
\newcommand{\dgzint} {\mbox{$[-0.12,\;\;0.11]$}} 
\newcommand{\dkgthd} {\mbox{$ 0.02^{+0.20}_{-0.15}$}}
\newcommand{\lamthd} {\mbox{$-0.190^{+0.087}_{-0.082}$}}
\newcommand{\dgzthd} {\mbox{$ 0.120^{+0.077}_{-0.083}$}}
\newcommand{\ddkgJT }    {\mbox{$ 0.040$}}
\newcommand{\ddkgBG }    {\mbox{$ 0.030$}}
\newcommand{\ddkgEF }    {\mbox{$ 0.038$}}
\newcommand{\ddkgEB }    {\mbox{$ 0.021$}}
\newcommand{\ddkgFR }    {\mbox{$ 0.008$}}
\newcommand{\ddkgMC }    {\mbox{$ 0.020$}}
\newcommand{\ddkgsu }    {\mbox{$ 0.070$}}
\newcommand{\dkgsta}    {\mbox{$ -0.12^{+0.28}_{-0.21}$}}
\newcommand{\dkgchi}    {\mbox{$ -0.11^{+0.30}_{-0.22}$}}
\newcommand{\dkgexp}    {\mbox{$\pm0.26$}}
\newcommand{\dkgiod}    {\mbox{$ -0.10^{+0.26}_{-0.21}$}}
\newcommand{\dkgbot}    {\mbox{$ -0.07^{+0.25}_{-0.19}$}}
\newcommand{\ddgzJT }    {\mbox{$ 0.005$}}
\newcommand{\ddgzBG }    {\mbox{$ 0.002$}}
\newcommand{\ddgzEF }    {\mbox{$ 0.010$}}
\newcommand{\ddgzEB }    {\mbox{$ 0.006$}}
\newcommand{\ddgzFR }    {\mbox{$ 0.012$}}
\newcommand{\ddgzMC }    {\mbox{$ 0.001$}}
\newcommand{\ddgzsu }    {\mbox{$ 0.018$}}
\newcommand{\dgzsta}    {\mbox{$ -0.029^{+0.074}_{-0.071}$}}
\newcommand{\dgzchi}    {\mbox{$ -0.034^{+0.077}_{-0.073}$}}
\newcommand{\dgzexp}    {\mbox{$\pm 0.068$}}
\newcommand{\dgziod}    {\mbox{$ -0.030^{+0.068}_{-0.066}$}}
\newcommand{\dgzbot}    {\mbox{$ -0.028^{+0.067}_{-0.065}$}}
\newcommand{\dlamJT }    {\mbox{$ 0.014$}}
\newcommand{\dlamBG }    {\mbox{$ 0.007$}}
\newcommand{\dlamEF }    {\mbox{$ 0.012$}}
\newcommand{\dlamEB }    {\mbox{$ 0.014$}}
\newcommand{\dlamFR }    {\mbox{$ 0.005$}}
\newcommand{\dlamMC }    {\mbox{$ 0.007$}}
\newcommand{\dlamsu }    {\mbox{$ 0.026$}}
\newcommand{\lamsta}    {\mbox{$ -0.129^{+0.075}_{-0.073}$}}
\newcommand{\lamchi}    {\mbox{$ -0.125^{+0.079}_{-0.076}$}}
\newcommand{\lamexp}    {\mbox{$\pm 0.073$}}
\newcommand{\lamiod}    {\mbox{$ -0.133^{+0.071}_{-0.068}$}}
\newcommand{\lambot}    {\mbox{$ -0.125^{+0.068}_{-0.065}$}}
\newcommand{\ddklJT }    {\mbox{$ 0.01$}}
\newcommand{\ddklBG }    {\mbox{$ 0.02$}}
\newcommand{\ddklEF }    {\mbox{$ 0.06$}}
\newcommand{\ddklEB }    {\mbox{$ 0.01$}}
\newcommand{\ddklMC }    {\mbox{$ 0.13$}}
\newcommand{\ddklsu }    {\mbox{$ 0.14$}}
\newcommand{\dklsta}    {\mbox{$ -0.10^{+1.32}_{-0.37}$}}
\newcommand{\dklchi}    {\mbox{$ -0.10^{+1.33}_{-0.37}$}}
\newcommand{\dklexp}    {\mbox{$\pm0.50$}}
\newcommand{\dkliod}    {\mbox{$ -0.17^{+0.53}_{-0.33}$}}
\newcommand{\dklbot}    {\mbox{$ -0.08^{+0.50}_{-0.31}$}}
\newcommand{\ddglJT }    {\mbox{$ - $}}
\newcommand{\ddglBG }    {\mbox{$ 0.02$}}
\newcommand{\ddglEF }    {\mbox{$ 0.02$}}
\newcommand{\ddglEB }    {\mbox{$ 0.01$}}
\newcommand{\ddglMC }    {\mbox{$ 0.03$}}
\newcommand{\ddglsu }    {\mbox{$ 0.05$}}
\newcommand{\dglsta}    {\mbox{$ -0.04^{+0.25}_{-0.35}$}}
\newcommand{\dglchi}    {\mbox{$ -0.04^{+0.25}_{-0.35}$}}
\newcommand{\dglexp}    {\mbox{$\pm 0.27$}}
\newcommand{\dgliod}    {\mbox{$ -0.17^{+0.28}_{-0.39}$}}
\newcommand{\dglbot}    {\mbox{$ -0.09^{+0.21}_{-0.20}$}}
\newcommand{\dlalJT }    {\mbox{$ - $}}
\newcommand{\dlalBG }    {\mbox{$ 0.02$}}
\newcommand{\dlalEF }    {\mbox{$ 0.02$}}
\newcommand{\dlalEB }    {\mbox{$ 0.01$}}
\newcommand{\dlalMC }    {\mbox{$ 0.02$}}
\newcommand{\dlalsu }    {\mbox{$ 0.04$}}
\newcommand{\lalsta}    {\mbox{$ -0.19\pm 0.15$}}
\newcommand{\lalchi}    {\mbox{$ -0.19\pm 0.15$}}
\newcommand{\lalexp}    {\mbox{$\pm 0.17$}}
\newcommand{\laliod}    {\mbox{$ -0.20\pm 0.14$}}
\newcommand{\lalbot}    {\mbox{$ -0.17^{+0.13}_{-0.12}$}}
\newcommand{\qqksta}   {\mbox{$0.36^{+0.68}_{-0.46}$}}
\newcommand{\qqgsta}   {\mbox{$0.08^{+0.16}_{-0.14}$}}
\newcommand{\qqlsta}   {\mbox{$0.05^{+0.21}_{-0.16}$}}
\newcommand{\qqktot} {\mbox{$ 0.31^{+0.73}_{-0.51}$}}
\newcommand{\qqgtot} {\mbox{$ 0.05^{+0.18}_{-0.15}$}}
\newcommand{\qqltot} {\mbox{$ 0.06^{+0.30}_{-0.20}$}}
\newcommand{\qqkexp} {\mbox{$\pm 0.57$}}
\newcommand{\qqgexp} {\mbox{$\pm 0.15$}}
\newcommand{\qqlexp} {\mbox{$\pm 0.19$}}
\newcommand{\qqkiod} {\mbox{$ 0.43^{+0.51}_{-0.43}$}}
\newcommand{\qqgiod} {\mbox{$ 0.12^{+0.17}_{-0.14}$}}
\newcommand{\qqliod} {\mbox{$ 0.17^{+0.31}_{-0.20}$}}
\newcommand{\qqkbot} {\mbox{$ 0.85^{+0.43}_{-1.06}$}}
\newcommand{\qqgbot} {\mbox{$ 0.16^{+0.17}_{-0.18}$}}
\newcommand{\qqlbot} {\mbox{$ 0.30^{+0.19}_{-0.30}$}}
\newcommand{\LepII}{\mbox{LEP}}
\newcommand{\LepI}{\mbox{LEP}}
\newcommand{\Jetset}{\mbox{J{\sc etset}}}
\newcommand{\Pythia}{\mbox{P{\sc ythia}}}
\newcommand{\Herwig}{\mbox{H{\sc erwig}}}
\newcommand{\Ariadne}{\mbox{A{\sc riadne}}}
\newcommand{\Excalibur}{\mbox{E{\sc xcalibur}}}
\newcommand{\Gentle}{\mbox{G{\sc entle}}}
\newcommand{\Racoon}{\mbox{R{\sc acoon}WW}}
\newcommand{\YFSww}{\mbox{Y{\sc fs}WW3}}
\newcommand{\Koralw}{\mbox{K{\sc oralw}}}
\newcommand{\Koralz}{\mbox{K{\sc oralz}}}
\newcommand{\Bhwide}{\mbox{B{\sc hwide}}}
\newcommand{\Phojet}{\mbox{P{\sc hojet}}}

\newcommand{\Com}{centre-of-mass}
\newcommand{\Grace}{\mbox{\tt grc4f}}
\newcommand{\SM}{Standard Model}
\newcommand{\MC}{Monte Carlo}
\newcommand{\tgc}{{\small TGC}}

\newcommand{\GeV}{\mbox{$\mathrm{GeV}$}}

\newcommand{\GeVcc}{\mbox{$\mathrm{GeV}\!/\!{\it c}^2$}}
\newcommand{\Ipb}{\mbox{pb$^{-1}$}}
\newcommand{\beq}{\begin{equation}}
\newcommand{\eeq}{\end{equation}}
\newcommand{\bea}{\begin{eqnarray}}
\newcommand{\eea}{\end{eqnarray}}
\newcommand{\ra}{\mbox{$\rightarrow$}}

\def\gappeq{\mathrel{ \rlap{\raise.5ex\hbox{$>$}}
                      {\lower.5ex\hbox{$\sim$}}  } }
\def\lappeq{\mathrel{ \rlap{\raise.5ex\hbox{$<$}}
                      {\lower.5ex\hbox{$\sim$}}  } }
\newcommand{\Mw}{\mbox{$M_{\mathrm{W}}$}}

\newcommand{\twsq}{\mbox{$\tan^2\theta_w$}}

\newcommand{\epem}{\mbox{$\mathrm{e^+e^-}$}}

\newcommand{\Zz}{\mbox{${\mathrm{Z}^0}$}}
\newcommand{\WW}{\mbox{$\mathrm{W^+W^-}$}}
\newcommand{\Wm}{\mbox{$\mathrm{W^-}$}}
\newcommand{\Wp}{\mbox{$\mathrm{W^+}$}}

\newcommand{\qq}{\mbox{$\mathrm{q\overline{q}}$}}

\newcommand{\qqqq}{\mbox{$\qq\qq$}}
\newcommand{\Wqq}{\mbox{$\mathrm{q\overline{q} }$}}

\newcommand{\ff}{\mbox{$\mathrm{f\overline{f}}$}}
\newcommand{\nnff}{\mbox{$\nu_{e}\overline{\nu}_{e}\ff$}}
\newcommand{\lnu}{\mbox{$\ell\overline{\nu}_{\ell}$}}
\newcommand{\lnubar}{\mbox{$\overline{\ell}^\prime\nu_{\ell^\prime}$}}

\newcommand{\enu}{\mbox{$\mathrm{e\overline{\nu}_{e}}$}}
\newcommand{\mnu}{\mbox{$\mu\overline{\nu}_{\mu}$}}
\newcommand{\tnu}{\mbox{$\tau\overline{\nu}_{\tau}$}}

\newcommand{\WWqqqq}{\mbox{\WW$\rightarrow$ \Wqq\Wqq}}
\newcommand{\WWqqln}{\mbox{\WW$\rightarrow$ \Wqq\lnu}}

\newcommand{\WWlnln}{\mbox{\WW$\rightarrow$ \lnu\lnubar}}

\newcommand{\Wenu}{\mbox{$\epem \rightarrow \mathrm{W}\enu$}}

\newcommand{\Zee}{\mbox{$\epem\rightarrow\Zz\epem$}}
\newcommand{\ZZ}{\mbox{$\epem\rightarrow\Zz\Zz$}}

\newcommand{\ZGqq}{\mbox{$\Zz/\gamma\rightarrow\qq$}}

\newcommand{\ZGff}{\mbox{$\Zz/\gamma\rightarrow\ff$}}

\newcommand{\Ebeam}{\mbox{$E_{\mathrm{beam}}$}}

\newcommand{\roots}{\mbox{$\sqrt{s}$}}

\newcommand{\Zgamma}{\mbox{$\Zz/\gamma$}}

\newcommand{\al}{\mbox{$\alpha$}}
\newcommand {\ee} {\mbox{$\mathrm{e}^+ \mathrm{e}^-$}}

\newcommand {\mm} {\mbox{$\mu^+ \mu^-$}}
\newcommand {\nn} {\mbox{$\nu\overline{\nu}$}}
\newcommand {\tautau} {\mbox{$\tau^+ \tau^-$}}

\newcommand {\eeee}   {\ee\ra\ee}
\newcommand {\eemumu} {\ee\ra\mm}
\newcommand {\eetautau} {\ee\ra\tautau}
\newcommand {\eenunu} {\ee\ra\nn}

\newcommand{\Cthw}{\mbox{$\cos\theta_{\mathrm{W}} $}}

\newcommand{\Cthstl}{\mbox{$\cos \theta_\ell^{*}$}}
\newcommand{\Phistl}{\mbox{$\phi_\ell^{*}$}}
\newcommand{\Cthstj}{\mbox{$\cos\theta_{\scriptscriptstyle \mathrm{jet}}^{*}$}}
\newcommand{\Phistj}{\mbox{$\phi_{\scriptscriptstyle \mathrm{jet}}^{*}$}}

\newcommand{\Cthsto}{\mbox{$\cos \theta_1^{*}$}}
\newcommand{\Phisto}{\mbox{$\phi_1^{*}$}}
\newcommand{\Cthstt}{\mbox{$\cos \theta_2^{*}$}}
\newcommand{\Phistt}{\mbox{$\phi_2^{*}$}}
\newcommand{\OO}{\mbox{${\cal O}$}}

\newcommand{\Lnln}{\lnu\lnubar}
\newcommand{\Qqln}{\Wqq\lnu}
\newcommand{\Qqen}{\Wqq\enu}
\newcommand{\Qqmn}{\Wqq\mnu}
\newcommand{\Qqtn}{\Wqq\tnu}
\newcommand{\Qqqq}{\Wqq\Wqq}

\newcommand{\gz}{\mbox{$g_1^{\mathrm{z}}$}}
\newcommand{\kg}{\mbox{$\kappa_\gamma$}}
\newcommand{\kz}{\mbox{$\kappa_{\mathrm{z}}$}}
\newcommand{\lgg}{\mbox{$\lambda_\gamma$}}
\renewcommand{\lg}{\mbox{$\lambda$}}
\newcommand{\lz}{\mbox{$\lambda_{\mathrm{z}}$}}
\newcommand{\dgz}{\mbox{$\Delta g_1^{\mathrm{z}}$}}

\newcommand{\dkg}{\mbox{$\Delta \kappa_\gamma$}}

\newcommand{\dkz}{\mbox{$\Delta \kappa_{\mathrm{z}}$}}

\newcommand{\LL}{\mbox{$\log L$}}

\newcommand {\xse}{cross-section}
\newcommand {\xses}{cross-sections}

\newcommand{\MOO}{\mbox{$\overline{\cal O}$}}
\newcommand{\OOEXP}[2]{\mbox{$E[{\cal O}_{#1}^{#2}]$}}

\documentstyle[epsfig,a4p,cite,12pt]{article}
\begin{document}
\bibliographystyle{plain}
\begin{titlepage}
\begin{center}{\large   EUROPEAN ORGANIZATION FOR NUCLEAR RESEARCH
 }\end{center}\bigskip
\begin{flushright}
% OPAL Physics Note PN425\\
CERN-EP/2000-114 \\ 
August 22, 2000  \\
%\today  \\
%Final Draft \\
\end{flushright}
\bigskip\bigskip\bigskip\bigskip\bigskip
\begin{center}
 {\huge\bf \boldmath 
Measurement of triple gauge boson couplings from \WW\ production 
at LEP energies up to 189 GeV}
\end{center}
\bigskip\bigskip
\begin{center}{\LARGE The OPAL Collaboration}
\end{center}\bigskip\bigskip
\bigskip\begin{center}{\large  Abstract}\end{center}
A measurement of triple gauge boson couplings is presented, based
on W-pair data recorded by the OPAL detector at LEP during 1998 at 
a centre-of-mass energy of 189 GeV with an integrated luminosity of 183 \Ipb. 
After combining with our previous measurements at centre-of-mass
energies of 161--183 GeV 
%including the single W process, 
we obtain
\kg=\reskg, \gz=\resgz\ and \lg=\reslam, where the errors
include both statistical and systematic uncertainties and each coupling
is determined by setting the other two couplings to their \SM\ 
values. These results are consistent with the Standard Model expectations. 
\bigskip\bigskip\bigskip\bigskip
%\bigskip\bigskip
%\begin{center}
%{\bf \large
%Principal authors: The \tgc\ working group \\
%Editorial board: E. Barberio, C.M. Hawkes, S. Spagnolo, M. Verzocchi \\
%Comments to Gideon.Bella@cern.ch  \\
%by Monday 21st August 2000. 
%}
%\end{center}
%{\large
%\begin{center}
% {\Large
%   This note describes preliminary OPAL results.}
%\end{center}
\begin{center}
{\large (Submitted to the European Physical Journal C)}
\end{center}

\end{titlepage}

%%%%%%%%%%%%%%%%%%%%%%%%%%%%%%%%%%%%%%%%%%%%%%%%%%%%%%%%%%%%%%%%%%%%%%%%%

\begin{center}{\Large        The OPAL Collaboration
}\end{center}\bigskip
\begin{center}{
%begin authorlist PLEASE DO NOT DELETE THIS COMMENT
G.\thinspace Abbiendi$^{  2}$,
K.\thinspace Ackerstaff$^{  8}$,
C.\thinspace Ainsley$^{  5}$,
P.F.\thinspace {\AA}kesson$^{  3}$,
G.\thinspace Alexander$^{ 22}$,
J.\thinspace Allison$^{ 16}$,
K.J.\thinspace Anderson$^{  9}$,
S.\thinspace Arcelli$^{ 17}$,
S.\thinspace Asai$^{ 23}$,
S.F.\thinspace Ashby$^{  1}$
% ...
%end authorlist PLEASE DO NOT DELETE THIS COMMENT
}\end{center}\bigskip
\bigskip
%begin institutes
$^{  1}$School of Physics and Astronomy, University of Birmingham,
Birmingham B15 2TT, UK
\newline
$^{  2}$Dipartimento di Fisica dell' Universit\`a di Bologna and INFN,
I-40126 Bologna, Italy
\newline
$^{  3}$Physikalisches Institut, Universit\"at Bonn,
D-53115 Bonn, Germany
\newline
% ...
$^{  5}$Cavendish Laboratory, Cambridge CB3 0HE, UK
\newline
% ...
$^{  8}$CERN, European Organisation for Nuclear Research,
CH-1211 Geneva 23, Switzerland
\newline
$^{  9}$Enrico Fermi Institute and Department of Physics,
University of Chicago, Chicago IL 60637, USA
\newline
% ...
$^{ 16}$Department of Physics, Schuster Laboratory, The University,
Manchester M13 9PL, UK
\newline
$^{ 17}$Department of Physics, University of Maryland,
College Park, MD 20742, USA
\newline
% ...
$^{ 22}$Department of Physics and Astronomy, Tel Aviv University,
Tel Aviv 69978, Israel
\newline
$^{ 23}$International Centre for Elementary Particle Physics and
Department of Physics, University of Tokyo, Tokyo 113-0033, and
Kobe University, Kobe 657-8501, Japan
\newline
% ...
%end notes
\newpage

%%%%%%%%%%%%%%%%%%%%%%%%%%%%%%%%%%%%%%%%%%%%%%%%%%%%%%%%%%%%%%%%%%%%%%%%

\section{Introduction}
 \label{sec:intro}

W-pair production in \epem\ annihilation involves,
in addition to the $t$-channel $\nu$-exchange, the triple gauge 
boson vertices WW$\gamma$ and WWZ which are present in the \SM\
due to its non-Abelian nature. Since the start of LEP operation
at and above the W-pair threshold several measurements have been made
of the triple gauge boson couplings (\tgc s) involved with these
vertices in W-pair production~\cite{tgc161-analysis,tgc172-analysis,
tgc183-analysis,OTHERLEPWW-tgc}, single W and single
photon production~\cite{OTHERpr}.
Limits on \tgc s also exist from studies of di-boson production at the
Tevatron~\cite{Tevatron-tgc}. In this paper we update our previous 
measurements~\cite{tgc161-analysis,tgc172-analysis,tgc183-analysis}
to include the data collected during 1998 at a centre-of-mass energy of 
189~GeV with an integrated luminosity of $\intLdt\,\Ipb$.

According to the most general Lorentz invariant 
Lagrangian~\cite{LEP2YR,HAGIWARA,BILENKY,GAEMERS} there can be
seven independent couplings describing each of the WW$\gamma$ and WWZ
vertices.
This large parameter space can be reduced by 
requiring the Lagrangian to satisfy electromagnetic gauge invariance
and charge conjugation as well as parity invariance. 
The number of parameters reduces to five, which can be taken
as \gz, \kz, \kg, \lz\ and \lgg~\cite{LEP2YR,HAGIWARA}.
These parameters are directly related to the W electromagnetic 
and weak properties~\cite{HAGIWARA,BILENKY}. 
In the \SM\, \gz=\kz=\kg=1 and \lz=\lgg=0.
Precision measurements on the \Zz\ resonance
and lower energy data, are consistent with the following
SU(2)$\times$U(1) relations 
between the five couplings~\cite{LEP2YR,BILENKY},
\begin{eqnarray*}
\label{su2u1}
\dkz & = & -\dkg \twsq +  \dgz, \\ 
\lz  & = &  \lgg. 
\end{eqnarray*} 
Here $\Delta$ indicates a deviation of the respective quantity 
from its \SM\, value and $\theta_w$ is the weak mixing angle.
These two relations leave only three independent couplings,
$\dkg$, $\dgz$ and \lg(=\lgg=\lz) which are not significantly 
restricted~\cite{DERUJULA,HISZ} by existing \LepI\ and SLC \Zz\ data.

Anomalous \tgc s give different contributions to different helicity
states of the outgoing W-bosons. Consequently they affect 
the angular distributions of the produced
W bosons and their decay products, as well as the total W-pair
\xse. Ignoring the effects of the finite
W width and the initial state radiation (ISR), the production
and decay of W bosons is completely described by five angles.
Conventionally these are taken to be~\cite{LEP2YR,BILENKY,SEKULIN}:
the \Wm\ production polar angle\footnote{
The OPAL right-handed coordinate system is defined 
such that the origin is at the geometric centre of the detector,
the $z$-axis is parallel to, and has positive sense, along the e$^-$ beam 
direction, $\theta$ is the polar angle with respect to $z$ and $\phi$ is 
the azimuthal angle around $z$.} $\theta_W$;
the polar and azimuthal angles, $\theta_1^*$ and $\phi_1^*$,
of the decay fermion from the \Wm\ in the \Wm\ rest frame\footnote{
The axes of the right-handed coordinate system in the W 
 rest-frame are defined such that $z$ is along
 the parent W flight direction and $y$ is in the direction 
 $\overrightarrow{e^-} \times \overrightarrow{W}$ where 
$\overrightarrow{e^-}$ is the electron beam 
 direction and $\overrightarrow{W}$ is the parent W flight direction.};
and the analogous angles for the anti-fermion from the \Wp\ 
decay, $\theta_2^*$ and $\phi_2^*$. However for W-pairs observed in the
detector, the experimental accessibility of these angles
and therefore the sensitivity to the \tgc s, is limited and depends 
strongly on the final state produced when the W bosons decay.

In this study all final states are used, namely 
the fully leptonic, \lnu\lnubar, the
semileptonic, \Wqq\lnu, and the fully hadronic, \Wqq\Wqq\ final states,
with branching fractions of 10.6\%, 43.9\% and 45.6\% respectively.
This study is divided into two parts. In the 
first part the W-pair event rate for each of the three final states
is analysed in terms of the \tgc s.
The second part is a study of the W-pair event shape for each final
state using optimal observables.

Most of the \tgc\ results of this analysis are obtained for each of the 
three couplings separately, setting the other two couplings to their
\SM\ values. These results will be presented in tables and 
\LL\ curves\footnote{
Throughout this paper, \LL\ denotes negative log-likelihood.}.
We also perform two-dimensional and three-dimensional fits, where
two or all three \tgc\ parameters are allowed to vary in the fits. 
The results of these fits will be presented by contour plots.  

The following section includes a short presentation of the OPAL data and 
Monte Carlo samples and in section \ref{sec:rate} the analysis of the W-pair 
event rate is described. Moving on to the event shape analysis, the 
first step, namely the event reconstruction, is presented in section 
\ref{sec:recon}. The \tgc\ extraction using the technique of optimal
observables is explained in section \ref{sec:oo}. 
Systematic errors are described 
in section \ref{sec:sys} and the combined \tgc\ results are presented 
in section \ref{sec:combtgc}. Section \ref{sec:summary} summarises the 
results of this study. 

%%%%%%%%%%%%%%%%%%%%%%%%%%%%%%%%%%%%%%%%%%%%%%%%%%%%%%%%%%%%%%%%%%%%%%%%%

\section{Data and Monte Carlo models}
\label{sec:data}
The data were acquired during 1998 with the OPAL detector which
is described in detail elsewhere~\cite{OPAL}. The integrated luminosity,
evaluated 
using small angle Bhabha scattering events observed in the silicon
tungsten forward calorimeter, is \intLdtfull\ $\pm$ \dLtot~\Ipb.
The luminosity-weighted mean centre-of-mass energy for the data sample
is \roots=\rroots$\pm$\droots~\GeV. 

In the analyses described below, a number of Monte Carlo models are
used to provide estimates of efficiencies and backgrounds as well as
the expected W-pair production and decay angular distributions for different 
\tgc\ values. The majority of the Monte Carlo samples were
generated at \roots\ = 189~GeV with $\Mw=80.33$~\GeVcc.  All Monte
Carlo samples mentioned below
were processed by the full OPAL simulation program~\cite{GOPAL} 
and then subjected to the same reconstruction procedure as the data.

The main \MC\ program used in this analysis is the 
\Excalibur~\cite{EXCALIBUR} generator. This program 
generates all four-fermion final states using the full set of
electroweak diagrams including the \WW\ production diagrams (class\footnote{
In this paper, the W pair production diagrams, {\em i.e.} 
$t$-channel $\nu_{\mathrm{e}}$ exchange and $s$-channel \Zgamma\ exchange, 
are referred to as ``CC03'', following the notation of \cite{LEP2YR}.} 
CC03) and other four-fermion graphs, such as \Wenu, \Zee\ and \ZZ.
Using this program, samples were generated with and without anomalous
\tgc s. Each sample was generated with a different set of \tgc\ values. 
These samples are used to calculate the \tgc\ dependence of the
expected event rate and angular distributions. We also apply a reweighting 
technique based on the 
matrix element corresponding to all the contributing four-fermion diagrams.
In this way, the angular distributions for any particular set of 
anomalous couplings are obtained from \MC\ samples generated at a
limited number of different \tgc\ values.

For systematic studies we use \SM\ four-fermion samples generated by 
the \Grace~\cite{GRC4F} and \Koralw~\cite{KORALW} programs.
\Koralw\ uses the same matrix element as \Grace\ but has the
most complete simulation of ISR out of all the four-fermion \MC\
programs used in this analysis. The efficiency obtained
from the \Koralw\ sample is used to correct the expected event rate.
To estimate the hadronisation systematics,
\MC\ samples were produced by \Grace\ including the CC03 diagrams only,
and the fragmentation stage was generated separately with either 
\Jetset~\cite{PYTHIA} or \Herwig~\cite{HERWIG}. To study final state 
interaction effects in \Qqqq\ events we used \MC\ samples with
different implementations of Bose-Einstein correlations and colour
reconnection effects, as detailed in section~\ref{sec:sys}.

There are other background sources not associated with four-fermion processes.
The main one, \ZGqq, including higher order QCD diagrams, is simulated 
using \Pythia, with \Herwig\ used as an alternative to study possible 
systematic effects. Other background processes involving two fermions 
in the final state are studied using \Koralz~\cite{KORALZ} for \eemumu, 
\eetautau\ and \eenunu, and \Bhwide\cite{BHWIDE} for \eeee. 
Backgrounds from two-photon processes are evaluated using \Pythia, 
\Herwig, \Phojet~\cite{PHOJET} and the Vermaseren 
generator\cite{VERMASEREN}. It is assumed that the centre-of-mass
energy of 189~\GeV\ is below the threshold for Higgs boson production.

%%%%%%%%%%%%%%%%%%%%%%%%%%%%%%%%% Event rate %%%%%%%%%%%%%%%%%%%%%%%%%%%%%

\section{Event rate TGC analysis}
\label{sec:rate}

The event rate analysis is based on the total \WW\ production \xse\
measurement~\cite{sigww189}. We use here the same selection with the 
same efficiency and background estimates. The difference is only in 
the physics interpretation.  
To analyse the \WW\ event rate, all three final states are used.
Each final state corresponds to a different selection algorithm. 
These algorithms are similar to those used at lower centre-of-mass
energies (see~\cite{tgc183-analysis} and references therein) and the
differences are described in~\cite{sigww189}.

The results are
summarised in Table~\ref{tab:rate}. The efficiencies refer to 
CC03 \WW\ events and include cross contamination between the three
final states. To calculate the expected numbers of events we use
our total integrated luminosity of $\intLdtfull\pm\dLtot$~\Ipb\ and
the total \WW\ production \xse\ value of 16.26 pb corresponding 
to our centre-of-mass 
energy of \rroots$\pm$\droots~GeV and a W mass of   
\Mw=80.42$\pm$0.07~\GeVcc\ measured at the Tevatron\footnote{
The \LepII\ results for the W mass are not used for the \tgc\ measurement, 
since they have been obtained under the assumption that W pairs are 
produced according to the \SM, whereas W production at 
the Tevatron does not involve the triple gauge vertex.}~\cite{MWPDG}.
The \xse\ calculation is done using the \Racoon~\cite{RacoonWW} and
\YFSww~\cite{YFSWW} \MC\ programs which include the most
complete ${\cal O}(\alpha)$ radiative corrections using the double pole
approximation method. The results obtained by these two independent 
programs agree within 0.1\%~\cite{FFMCWG}. Furthermore the theoretical 
uncertainty of 0.42\%
is improved compared with the 2\% uncertainty of the \Gentle~\cite{GENTLE} 
semi-analytic program which has been used in our previous publications.
However, since we have not yet been able to study the full effect 
of our selection cuts with these recent \MC\ generators, we prefer
not to reduce our assumed theoretical uncertainty of 2\% which
has only a negligible effect on our final results. 
\begin{table}[htbp]
 \begin{center}
 \begin{tabular}{|l|c|c|c|c|} \hline
Selected as     & \WWlnln & \WWqqln & \WWqqqq & Combined \\ \hline
Efficiency [\%] & 82.3$\pm$1.2 & 87.7$\pm$0.9 & 86.5$\pm$0.8 & 
                                              86.6$\pm$0.6 \\ \hline
Signal events   & 258$\pm$6 & 1145$\pm$26 & 1175$\pm$26 & 
                                              2578$\pm$55 \\ \hline
\underline{Background events:} &        &        &         &          \\
4-fermion, \tgc -dep.  & 17$\pm$2 & 27$\pm$4 & 10$\pm$6 &
                                                54$\pm$7 \\
4-fermion, \tgc -indep. & 7.0$\pm$0.3 & 34$\pm$3 & 71$\pm$10 & 
                                                111$\pm$9 \\
\ZGff           & 4.1$\pm$0.3 & 48$\pm$6 & 245$\pm$17 & 
                                               297$\pm$17  \\
Two-photon      & 0.9$\pm$0.9 & 3$\pm$3 &    0        & 
                                           4$\pm$4 \\  \hline     
Total background & 29$\pm$3 & 112$\pm$9 & 325$\pm$21 & 
                                        466$\pm$24 \\  \hline
Total expected  & 287$\pm$7 & 1257$\pm$27  & 1500$\pm$33 & 
                                         3044$\pm$60 \\   \hline
Observed        & 276 & 1246 & 1546 & 3068  \\ \hline 
 \end{tabular}
 \end{center}
\caption{Expected and observed numbers of events in each \WW\ final
 state for an integrated luminosity of 
 $\intLdtfull\pm\dLtot$~\Ipb\ at $\rroots\pm\droots$~\GeV, assuming 
  $\Mw=80.42\pm0.07$~\GeVcc\ and \SM\ branching fractions. 
  The efficiency values and the numbers of signal events 
  include cross contamination between the three final states. 
  The errors on the combined numbers account for 
  correlations between the systematic errors for different final states.}
\label{tab:rate}
\end{table}

The four-fermion background is split between final states with 
and without contributions from diagrams containing a triple gauge boson 
vertex. For each of the three final states in our signal
the \tgc -dependent background is calculated from the difference between 
the accepted four-fermion cross-section (including all diagrams)
and the accepted CC03 cross-section. The background from other final
states\footnote{
A small contribution from the \nnff\ final states which are partly
produced by the W fusion diagram is included in the 
\tgc -dependent background.} does not depend on the \tgc s.

The errors listed in Table~\ref{tab:rate} are due to \MC\ statistics, 
luminosity uncertainty of 0.2\%, theoretical uncertainty in the total 
\xse\ (2\%), centre-of-mass energy 
and W mass uncertainties (0.04\% of the total \xse\ each),
data/MC differences, tracking losses, detector
occupancy and fragmentation uncertainties. A detailed description of
all these sources can be found in~\cite{sigww189}.

The total number of expected events in each
final state is consistent with the corresponding number of observed events.
Therefore, there is no evidence for any significant contribution
from anomalous couplings. 
A quantitative study of \tgc s from the W-pair event yield is performed by 
comparing the numbers of observed events in each of the three event
selection channels with the expected numbers which are parametrised
as second-order polynomials in the \tgc s. This parametrisation is
based on the linear dependence of the triple gauge vertex Lagrangian
on the \tgc s, corresponding to a second-order polynomial dependence of 
the \xse. Consequently, the expected number of  
events for each final state also has a second-order
polynomial dependence on the \tgc s. The polynomial coefficients are 
calculated from the expected number of events at different \tgc\ values, 
as obtained from the corresponding \Excalibur\ \MC\ samples. For each final
state, the corresponding expected cross section is normalised to the
\SM\ expectation (sum of signal and \tgc -dependent background) listed in 
Table~\ref{tab:rate}. The primary reason for this normalisation is the 
fact that when calculating the numbers in Table~\ref{tab:rate}, \Racoon\ and 
\YFSww\ are used for the \xse\ and \Koralw\ for the efficiency. 
This is considered to be more complete than \Excalibur. 
The normalisation factor varies between 
0.970 (\Lnln\ selection) and 0.987 (\Qqqq\ selection).  
   
The probability to observe the measured number of candidates, given the 
expected value, is calculated using a Poisson distribution.
The product of the three probability distributions 
corresponding to the three final states is taken as the event rate
likelihood function. 
The systematic uncertainties are incorporated into the \tgc\ fit by
allowing the expected numbers of signal and background events to 
vary in the fit and constraining them to have a Gaussian distribution
around their expected values with their systematic errors taken as the
width of the distributions. The systematic uncertainties, excluding 
those on efficiency and background, are 
assumed to be correlated between the three different event selections.
 
\begin{figure}[htbp]
\epsfig{file=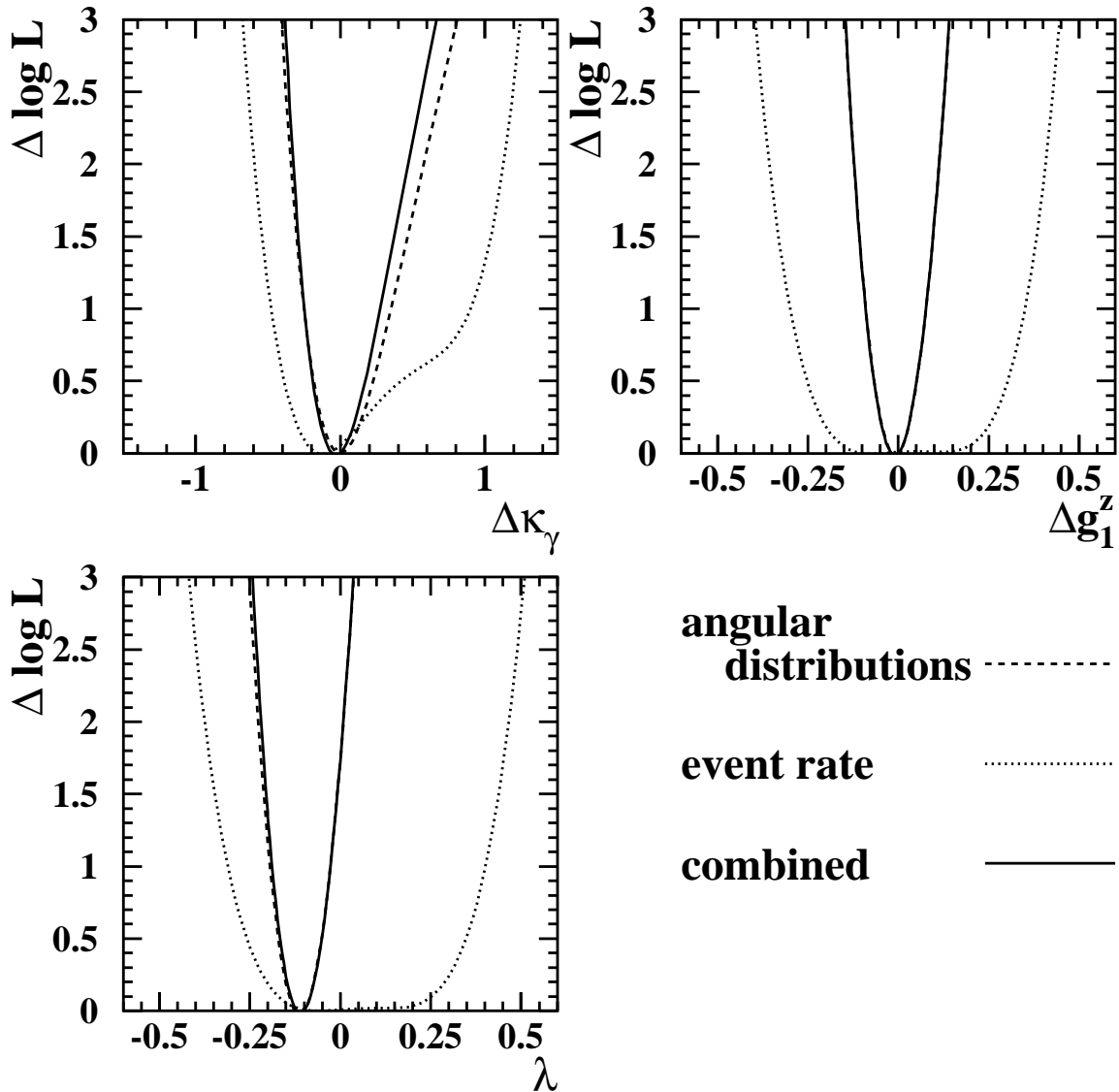,width=\textwidth}
\caption{\sl
Negative log-likelihood curves obtained using different sources
of information on the TGCs.
The curves for each TGC parameter are obtained
setting the other two parameters to their \SM\ values. The dashed lines 
are obtained from the event shapes and the dotted lines from the event rates. 
%and the dash-dotted lines are obtained from the single W analysis. 
The systematic errors are included. The solid lines are obtained by combining 
%the three
both 
sources of information. 
%Note that the single W analysis does not provide any information on \dgz.
}
\label{fig:rateshape}
\end{figure}

Data from lower centre-of-mass energies are included, assuming all
systematic errors to be fully correlated between energies. 
The corresponding \LL\ curves are used in combination with the
results of the event shape analysis, which is described
in the following sections. The full set of results is then presented
in Figure~\ref{fig:rateshape}, where the event rate contributions
are shown as dotted lines\footnote{
The \LL\ curves are expected to be 
symmetric around the event rate minimum position which, in most cases,
is very close to the \SM\ \tgc\ value. However, for the \dkg\ parameter,
the event rate acquires its minimum at positive \dkg\ value.
The exact location of this minimum differs between \lnu\lnubar\, 
\Wqq\lnu\ and \Wqq\Wqq\ events due to contributions from non-CC03
diagrams. Therefore, the overall \LL(\dkg) function summed over
the three final states is no longer symmetric, unlike the 
corresponding functions for the other \tgc\ parameters. 
}. 

%%%%%%%%%%%%%%%%%%%%%%%%%%%%%%%%%%%%%%%%%%%%%%%%%%%%%%%%%%%%%%%%%%%%%%%%%

\section{Event reconstruction}
\label{sec:recon}
Starting from the event sample used for the event-rate analysis,
a reconstruction is performed to extract the maximum possible information  
on the W production and decay angles, which are then used to 
extract the couplings. Events which cannot be well 
reconstructed are rejected from the sample.

  Three kinematic fits with different sets of requirements are used in 
the event reconstruction:
\begin{itemize}
\item[A.] requiring conservation of energy and momentum, neglecting ISR;
\item[B.] additionally constraining the reconstructed masses of the two 
W-bosons to be equal;
\item[C.] additionally constraining each reconstructed W mass to the 
%nominal value.
average value measured at the Tevatron, \Mw=80.42 \GeVcc~\cite{MWPDG}. 
\end{itemize}
For \qqqq\ events, where all four final state fermions are measurable, 
fits A, B and C have 4, 5 and 6 constraints respectively.
For \Qqen\ and \Qqmn\ events the number of constraints is reduced by 3 
due to the invisible neutrino. For \Qqtn\ events there 
is at least one additional unobserved neutrino from the $\tau$ decay,
but the momentum sum of the track(s) assigned to the $\tau$ can still be
used as an approximation to the $\tau$ flight direction, relying
on its high boost. The $\tau$ energy is left 
unknown, resulting in a further loss of one constraint. Finally for
\Lnln\ events, where none of the leptons is a $\tau$, there are two
invisible neutrinos. Hence, six constraints are lost and requirement C
is needed.

  In the following we discuss the reconstruction of each final state
separately.

%%%%%%%%%%%%%%%%%%%%%%%%%%%%%%%%%%%%%%%%%%%%%%%%%%%%%%%%%%%%%%%%%%%%%

\subsection{Reconstruction of \Qqln\ final states}
\label{subs:recqqln}
 
Candidate \Qqen\ and \Qqmn\ events without a well reconstructed lepton 
track, which were included in the sample used for the event rate 
analysis, are removed from the present sample. In this way, each of the 
events left in the sample has one track that is identified 
as the most likely lepton candidate.
For the case of \Qqtn\ events, either one track or a narrow jet 
consisting of three tracks is assigned as the $\tau$ decay product.

The electron direction is reconstructed  
by the tracking detectors and the energy is measured in 
the electromagnetic calorimeters. For muons the momentum is 
measured using the tracking detectors. As explained above,
the direction of $\tau$ candidates can be directly reconstructed
whilst the energy can only be derived from a kinematic fit.   

The remaining tracks and 
calorimeter clusters in the event are grouped into two jets
using the Durham $k_\perp$ algorithm~\cite{DURHAM}. The total 
energy and momentum of each of the jets are calculated with the method
described in~\cite{GCE}. Kinematic fit A, requiring energy-momentum
conservation, is then performed 
on the events. The \Qqen\ and \Qqmn\ events are accepted if this 
one-constraint fit converges with a probability larger than 0.001. 
This cut rejects about 2\% of the signal events and 4\% of the background.    

To improve the resolution in the angular observables used for the 
\tgc\ analysis, kinematic fit C is performed on \Qqen\ 
and \Qqmn\ events. The W-mass constraint in this fit allows for the
finite W-width\footnote{
 The W mass distribution is treated as a Gaussian in the kinematic fit. 
 However, in 
 order to simulate the expected Breit-Wigner form of the W mass spectrum,
 the variance of the Gaussian is updated at each iteration of the 
 kinematic fit in such a way that the probabilities of observing 
 the current fitted W mass are equal whether calculated
 using the Gaussian distribution or using a Breit-Wigner.}.
We demand that the kinematic fit converges with a probability larger than 
0.001. For the 4\% of events which fail at this point we revert
to using the results of fit A.

For \Qqtn\ events the kinematic fit B is performed and 
required to converge with a probability larger than 0.025. This cut 
rejects 14\% of the signal and 41\% of the background. Furthermore, 
this cut suppresses those \Qqtn\ events which are 
correctly identified as belonging to this decay channel 
but where the $\tau$ decay products are not identified correctly, 
leading to an incorrect estimate of the $\tau$ flight direction or 
its charge. The fraction of such events in the \Qqtn\ sample
is reduced from 18\% to 12\%.

Out of the original number of 1246 \Qqln\ candidates used for the event 
rate analysis there are 1075 events left
(368 \Qqen, 387 \Qqmn\ and 320 \Qqtn) after all these cuts. 
Assuming \SM\ \xses\ for signal and background processes,
the remaining background fraction
is 4.6\%. This does not include cross-migration between the three lepton
channels. The background sources are: four-fermion processes after 
subtracting the contribution from the CC03 diagrams (2.4\%), \ZGqq\ (2.4\%),
\WWqqqq\ and \Lnln\ (0.2\%) and two-photon reactions (0.1\%). 

In the reconstruction of the \Qqln\ events we obtain $\Cthw$ by summing 
the kinematically fitted four-momenta of the two jets. 
The decay angles of the leptonically decaying W are obtained from the 
charged lepton four-momentum, after boosting back to the parent W rest 
frame. For the hadronically decaying W we are left with a two-fold ambiguity 
in assigning the jets to the quark and antiquark. This ambiguity is 
taken into account in the analyses described below. 

\begin{figure}[tbhp]
 \epsfxsize=\textwidth
 \epsffile{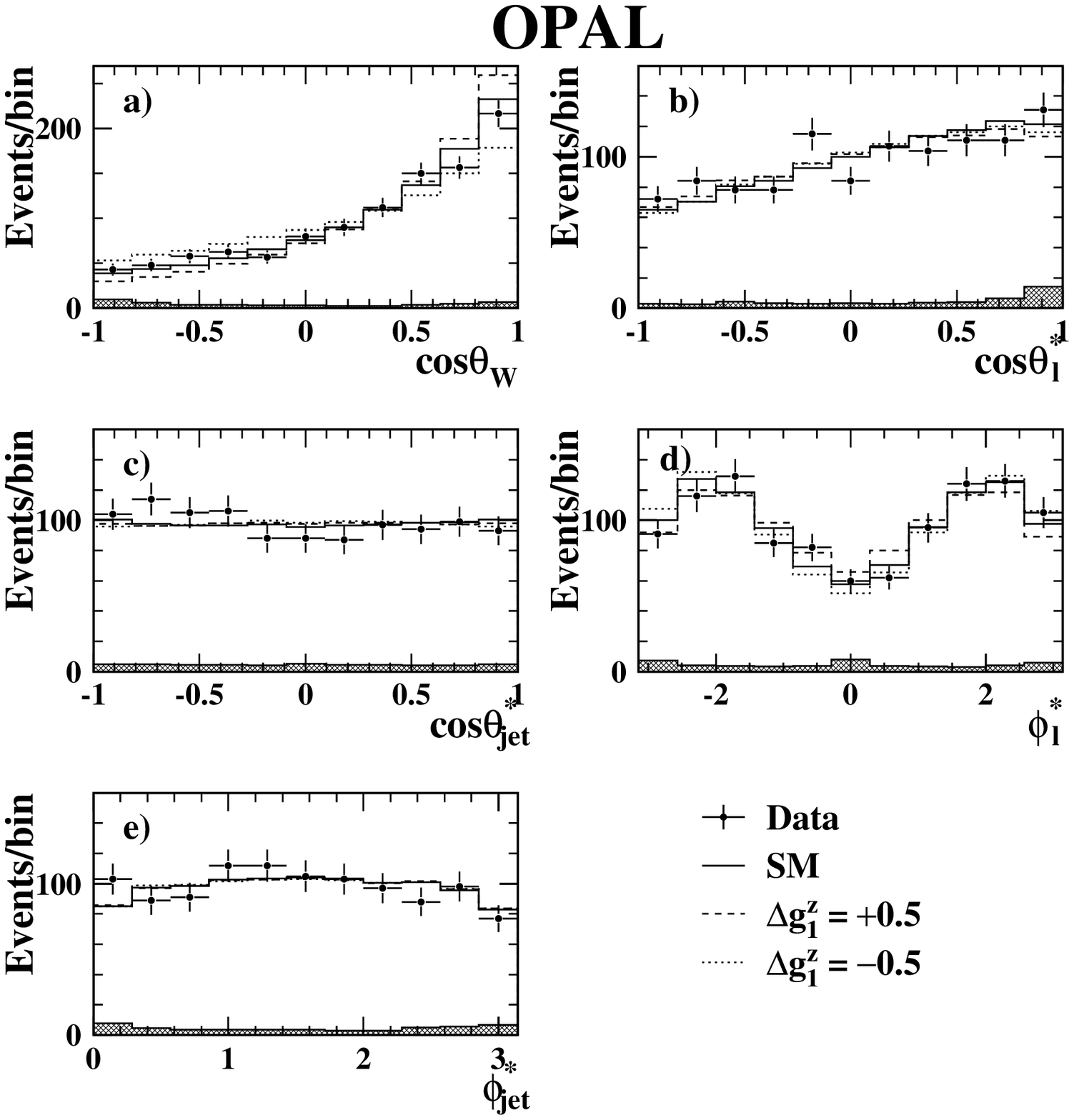}
   \caption{\sl
    Distributions of the kinematic variables \Cthw, \Cthstl,
    \Cthstj, \Phistl\ and \Phistj, 
    as obtained from the \Qqln\ events. The solid points
    represent the data. The histograms show the expectation of the 
    \SM\ and when $\dgz=\pm 0.5$. The shaded histogram shows the 
    non-\Qqln\ background.
    Notes: 1. In the case of W$^+\ra\bar{\ell}\nu$ 
    decays the value of $\Phistl$
    is shifted by $\pi$ in order to overlay W$^+$ and W$^-$ 
    distributions in the same plot.
    2. The jet with $0\leq\Phistj\leq\pi$ is chosen as the 
    quark (antiquark) jet from the decay of the \Wm\ (\Wp) according
    to our convention. }
 \label{fig:angdist} 
\end{figure}

In Figure~\ref{fig:angdist} we show the distributions of all the five
angles obtained from the combined \Qqln\ event sample,
 and the expected distributions for $\dgz = \pm 0.5$ and 0.
These expected distributions are obtained from fully simulated
\MC\ event samples generated with \Excalibur, normalised
to the number of events observed in the data. Sensitivity to \tgc s
is observed mainly for \Cthw. The contribution of \Cthstl, \Phistl, 
\Cthstj\ and \Phistj\ to the overall
sensitivity enters mainly through their correlations with \Cthw.
 
%%%%%%%%%%%%%%%%%%%%%%%%%%%%%%%%%%%%%%%%%%%%%%%%%%%%%%%%%%%%%%%%%%%%%

\subsection{Reconstruction of \Qqqq\ final states}
\label{subs:recqqqq}
 
Using the Durham $k_\perp$ algorithm \cite{DURHAM}, each selected 
\Qqqq\ candidate is reconstructed as four jets, whose energies 
are corrected for the double counting of charged track momenta and 
calorimeter energies~\cite{MT}. These four jets can be paired 
into W-bosons in three possible ways. 
To improve the resolution on the jet four-momentum we perform  
for each possible jet pairing the five-constraint kinematic fit B. 
We require at least one jet pairing with a successful fit yielding a
W-mass between 70 and 90 GeV. The W charge is assigned by comparing
the sum of the charges of the two jets coming from the same W candidate. 
More details are described in~\cite{tgc183-analysis}.

The correct di-jet pairing is identified
by a likelihood algorithm~\cite{Karlen} which 
takes into account correlations between the input variables. 
We use as input to our likelihood 
the W candidate mass as obtained from the kinematic fit B, 
the angles between the two jets coming from each W candidate, 
the difference between the W candidate masses 
as obtained from the kinematic fit A
and the charge difference between the two W candidates.
According to the \WW\ \Excalibur\ Monte Carlo, 
the probability of selecting the correct di-jet 
combination is about 79$\%$, and the 
probability of correct assignment of the W charge, once the correct
pairing has been chosen, is about 77$\%$. 

\begin{figure}[htbp]
  \begin{center}
    \leavevmode

\epsfig{file=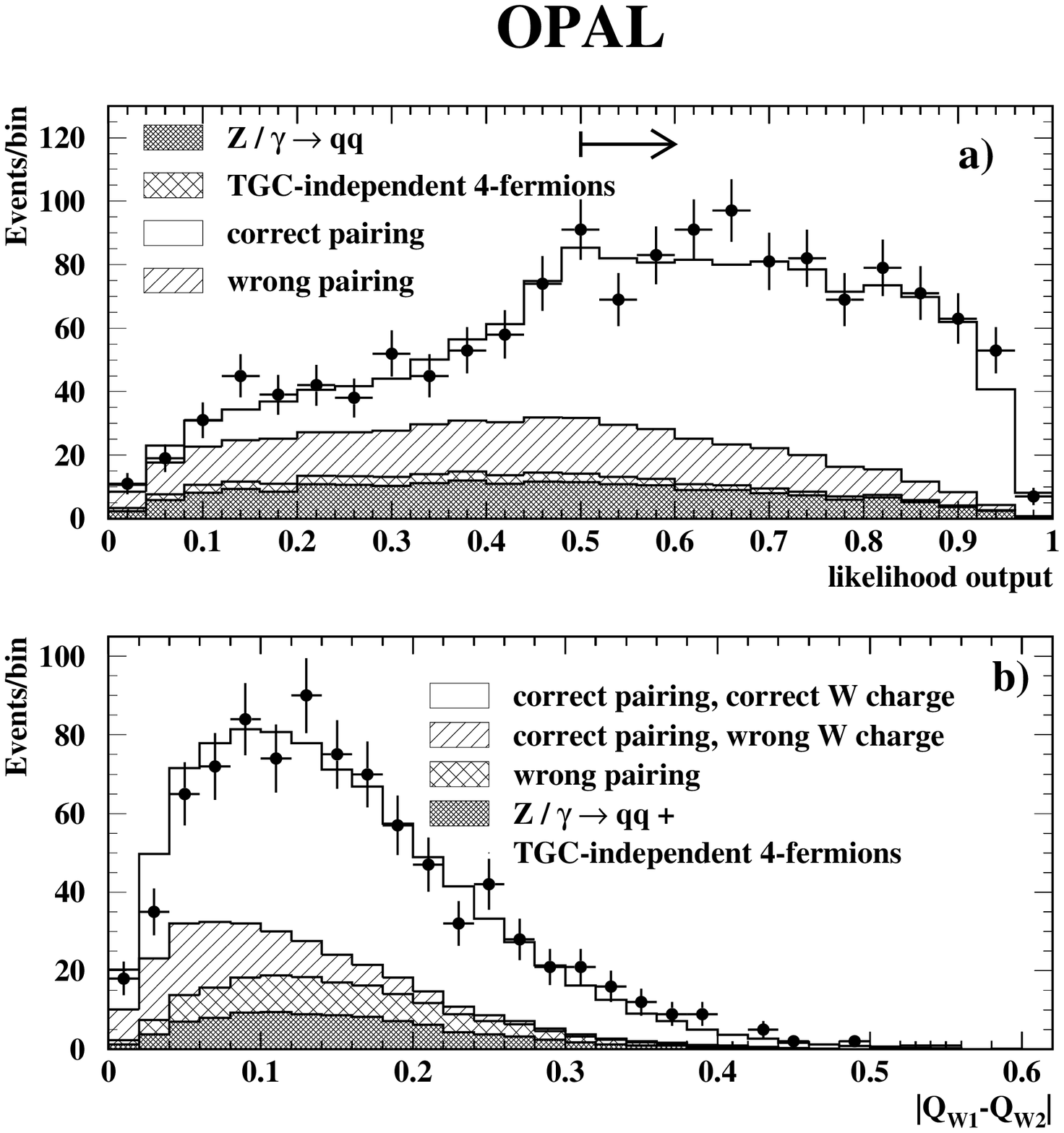,bbllx=24pt,bblly=160pt,bburx=530pt,bbury=694pt,
     width=\linewidth,clip=}
    \caption{\sl
     For the \qqqq\ channel, 
     (a) distribution of the output of the jet-pairing likelihood
     corresponding to the most likely combination 
     for the data (points)
     and for the Excalibur four-fermion Monte Carlo (histogram).
     The hatched area represents the contribution of
     wrong pairing (assuming all four quark final states 
     compatible with W pair production are coming
     from such events), the double hatched area
     represents the contribution of the TGC-independent four-fermion 
     background and the dark area represents
     the contribution of the \ZGqq\ background. 
     The arrow indicates the cut value.
     (b) Distribution of the charge separation between the two W
     candidates  for the data (points)
     and for the four-fermion Monte Carlo (histogram).
     The hatched area shows the
     contribution of correct pairing and incorrect W charge, 
     and the double hatched area shows the contribution
     of wrong pairing. The dark area represents
     the contribution  of the \ZGqq\ and TGC-independent four-fermion 
     background.}    
    \label{fig:likout_qw}
  \end{center}
\end{figure}

The fraction of events correctly
reconstructed can be increased by cutting on the value $L$ of the 
jet pairing likelihood. Its distribution is shown in 
Figure~\ref{fig:likout_qw}(a). We cut at $L > 0.5$,
which minimises the expected total errors on the
measured \tgc s.

\begin{figure}[htbp]
  \begin{center}
    \leavevmode

\epsfig{file=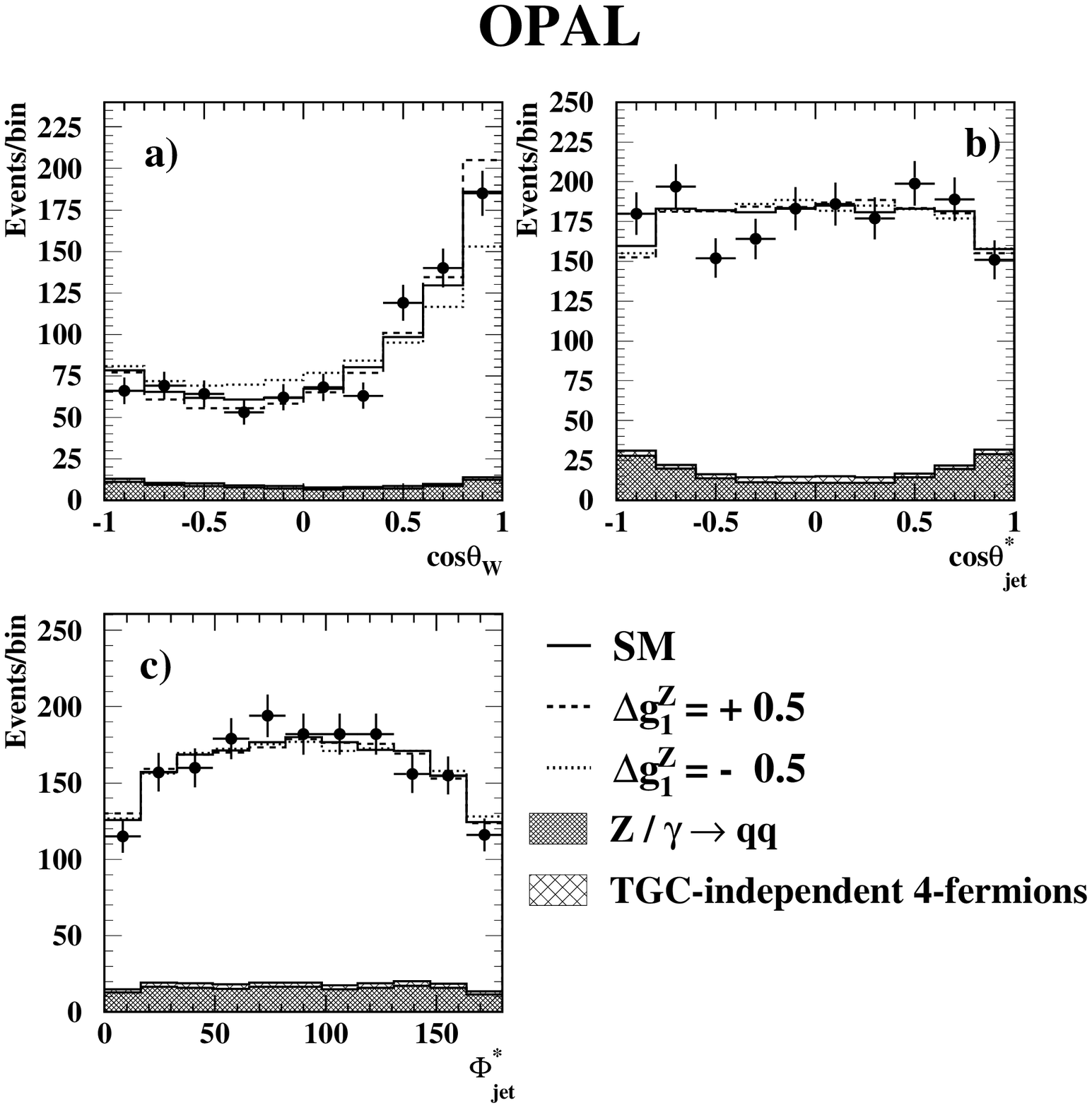,bbllx=24pt,bblly=150pt,bburx=560pt,bbury=694pt,
     width=\linewidth,clip=}
    \caption{\sl
     Distribution of the kinematic variables \Cthw, \Cthstj, 
     and \Phistj\ for \Qqqq\ events.
     The solid points represent the data. The Monte Carlo
     predictions for the Standard Model and for \dgz=+0.5, $-$0.5 
     are shown respectively as
     solid, dashed and dotted lines. The dark area represents the
     contribution of the \ZGqq\ background and the hatched area
     represents the contribution of the TGC-independent four-fermion 
     background. The jet with $0\le$\Phistj$\le\pi$ is 
     chosen as the quark (antiquark) jet from the decay of the
     W$^-$ (W$^+$) candidate according to our convention.} 
    \label{fig:kineqqqq}
  \end{center}
\end{figure}

After all cuts the probability of correct
pairing increases to 87\% and the
probability of correct determination of the W charge is about 82\%.
The selection efficiency is reduced by these cuts from 86\% to 54\%, 
whereas  the background drops from 23\% to 12\%. In the data,
889 events survive all cuts out of the 1546 initially selected.
The distribution of the W charge separation is shown in
Figure~\ref{fig:likout_qw}(b). The distributions of the variables 
\Cthw, \Cthstj\ and \Phistj\ are shown in Figure~\ref{fig:kineqqqq}. 

%%%%%%%%%%%%%%%%%%%%%%%%%%%%%%%%%%%%%%%%%%%%%%%%%%%%%%%%%%%%%%%%%%%%%

\subsection{Reconstruction of \Lnln\ final states}
\label{subs:reclnln}
 
As explained above, the reconstruction of \Lnln\ final states is 
possible only by using kinematic fit C for the case of no $\tau$-lepton in 
the final state. To reject events with $\tau$-leptons we use the lepton
identification algorithm designed for the W branching ratio
measurement~\cite{sigww189}. In addition, events with only one
reconstructed cone which were accepted for the event rate analysis
are removed from the sample. As a result of these two cuts, the 
number of data events drops from 276 to 133. The efficiency for \Lnln\ 
events with $\ell=$e,$\mu$ drops from 89\% to 74\%. The contamination
left in the sample from \Lnln\ final states with a $\tau$-lepton is
14\% and the background from other final states is only 0.5\%.

Kinematic fit C has no constraints and turns into solving a quadratic 
equation. In the ideal case of no measurement errors and satisfying the 
conditions where fit C is valid, namely no ISR and narrow W width, one 
expects to obtain two real solutions corresponding to a two-fold ambiguity 
in the angle set $\cos\theta_W$, $\phi_1^*$ and $\phi_2^*$. There is
no ambiguity in the angles $\theta_1^*$ and $\theta_2^*$ which have a
one-to-one correspondence with the lepton momenta. 
In realistic conditions, however, including ISR, finite W width and 
measurement errors, one might obtain no real solution, but a pair 
of complex conjugate solutions. In fact, only 92 out of 133 events in
the data have two real solutions, in agreement with the Monte Carlo 
expectation of 97.8 events. In the complex solution case, the nearest
real solution is taken by setting the imaginary parts of the complex 
solutions to zero.

\begin{figure}[htb]
\epsfig{file=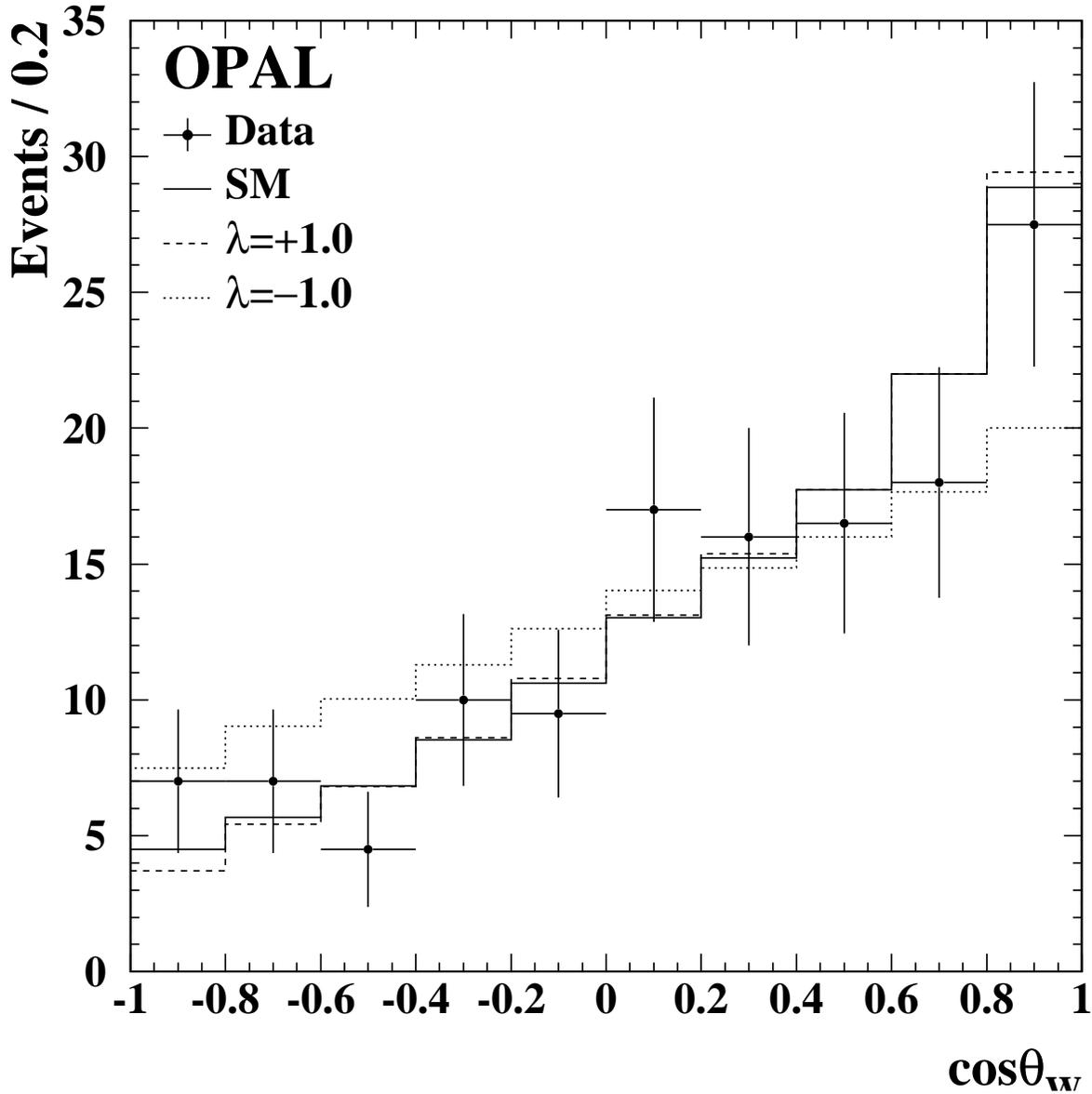,width=\textwidth}
\caption{\sl
Observed \Cthw\ distribution in the \Lnln\ analysis. All events enter
with a total weight of one, but in the case of events with two ambiguous
solutions for \Cthw, each solution enters with a weight of 0.5.
The histograms show the expectations of the \SM\ and when $\lg=\pm 1$.
}
\label{fig:dacostw}
\end{figure}

Figure~\ref{fig:dacostw} shows the reconstructed \Cthw\ distribution.
The data distribution agrees with the \SM\ expectation.
%Figure~\ref{fig:dacos12} shows the joint distribution of     
%$\cos\theta^*_1$ and $\cos\theta^*_2$. Particularly interesting is the  
%\lg-dependence of the correlation between the two $\theta^*_i$ angles.
%This effect cannot be seen in other W-pair final states because of 
%the difficulty in distinguishing the quark from the antiquark direction
%in hadronic W decays. 

%%%%%%%%%%%%%%%%%%%%%%%%%%%%%%%%%%%%%%%%%%%%%%%%%%%%%%%%%%%%%%%%%%%%%%%%%%

\section{Event shape TGC analysis}   
\label{sec:oo} 

The angular distributions from all final states are consistent with the 
\SM\ expectation and no evidence is seen for any significant
contributions from anomalous couplings. For a quantitative study we
use the method of optimal observables based on the following
dependence of the W-pair differential \xse\ on the \tgc s,
\[
d\sigma(\Omega,\al)=S^{(0)}(\Omega) + \sum_i \al_i \cdot S_i^{(1)}(\Omega)  + 
\sum_{i,j} \al_i \al_j \cdot S^{(2)}_{ij}(\Omega),\ \ \
   \al_i \ = \ \dkg,\ \dgz \  {\rm and} \  \lg,  
\]
where $\Omega$ are the five phase-space variables,
$\Omega$=(\Cthw,\Cthsto,\Phisto,\Cthstt,\Phistt).
For this kind of dependence it has been shown~\cite{Fanourakis} that
all information contained in the phase-space variables $\Omega$ is retained 
in the whole set of nine observables 
$\OO_i^{(1)}=S_i^{(1)}(\Omega)/S^{(0)}(\Omega)$,
$\OO_i^{(2)}=S_{ii}^{(2)}(\Omega)/S^{(0)}(\Omega)$ and
$\OO_{ij}^{(2)}=\OO_{ji}^{(2)}=S_{ij}^{(2)}(\Omega)/S^{(0)}(\Omega)$ 
$(i,j=1,2,3)$.

The original idea~\cite{Rouge,Diehl,Fanourakis2} was to use only the
mean values of the first order observables, $\MOO^{(1)}$. 
However, this causes some ambiguities which can be resolved by 
inclusion of the mean value of the second order observables,
$\MOO^{(2)}$. 
For one-dimensional fits, where only one \tgc\ parameter, $\al_i$, is 
allowed to vary and the others are set to their \SM\ values, we use 
only the corresponding first order, $\OO_i^{(1)}$, and second order,
$\OO_i^{(2)}$, observables. In three-dimensional fits, where all
three \tgc\ parameters are allowed to vary, all nine observables are 
used\footnote{ 
Trying to use all the nine observables also for the one-dimensional 
fits does not yield any gain in sensitivity, since the optimal 
observables not related to the fitted parameter are not expected 
to contribute further information.}.

%For one-dimensional fits, where only one \tgc\ parameter, $\al_i$, is 
%allowed to vary and the others are set to zero, it is sufficient to 
%consider only the corresponding first order, $\OO_i^{(1)}$, and 
%preferably also the second order, $\OO_i^{(2)}$, observable\footnote{
%Neglecting the contribution of the second order observable might cause
%a loss in sensitivity and a wrong estimate of the statistical errors.
%This can be partially corrected by using a complicated iterative
%procedure~\cite{KocianDiehl,Fanourakis}.}.
%However, in three-dimensional fits where all three \tgc\ parameters are
%allowed to vary, it would be best to use all nine observables.
  
%In our previous publication \cite{tgc183-analysis} we used this 
%optimal observable method only for one-dimensional fits and 
%analysed the distributions of the first order observables only.
%Extending this procedure to three-dimensional fits would require
%handling the joint distribution of all nine observables. In this 
%case it would be better to abandon the use of optimal observables 
%and revert to the original five phase-space variables $\Omega$.

%In order to simplify the method we revert to the 
%original  idea~\cite{Rouge,Diehl,Fanourakis2} of using only the mean 
%values of the first order observable $\MOO^{(1)}$, and we extend 
%this method by inclusion of the mean values of the second order
%observables, $\MOO^{(2)}$.

The analysis is done separately for the \Qqln, \Qqqq\ and 
\Lnln\ final states. The nine optimal observables 
are constructed for each event $k$ with the set of phase-space variables 
$\Omega_k$ using the analytic expression for the CC03 Born differential 
\xse\ to calculate the values 
of $S^{(0)}(\Omega_k), \ S_i^{(1)}(\Omega_k), \ S_{ii}^{(2)}(\Omega_k)$ 
and $S_{ij}^{(2)}(\Omega_k)$, 
taking into account the reconstruction ambiguities for that particular
final state. For the \Qqqq\ channel, the uncertainty in the identity 
of the W$^+$ and the W$^-$ is taken into account by 
weighting each hypothesis with its probability, determined
from the Monte Carlo
as a function of the charge difference $|Q_{W_1}-Q_{W_2}|$
between the two W candidates in the event.

The expected mean values of each observable as a function 
of the \tgc s are calculated from four-fermion \MC\ events generated 
according to the \SM\ as well as samples generated with various \tgc\ 
values. The \Excalibur\ matrix element calculation is used 
to reweight the \MC\ events to any \tgc\ value required. The contribution 
of background events is also taken into account in the calculation of 
the expected mean values using corresponding \MC\ samples.

\begin{figure}[htbp]
\epsfig{file=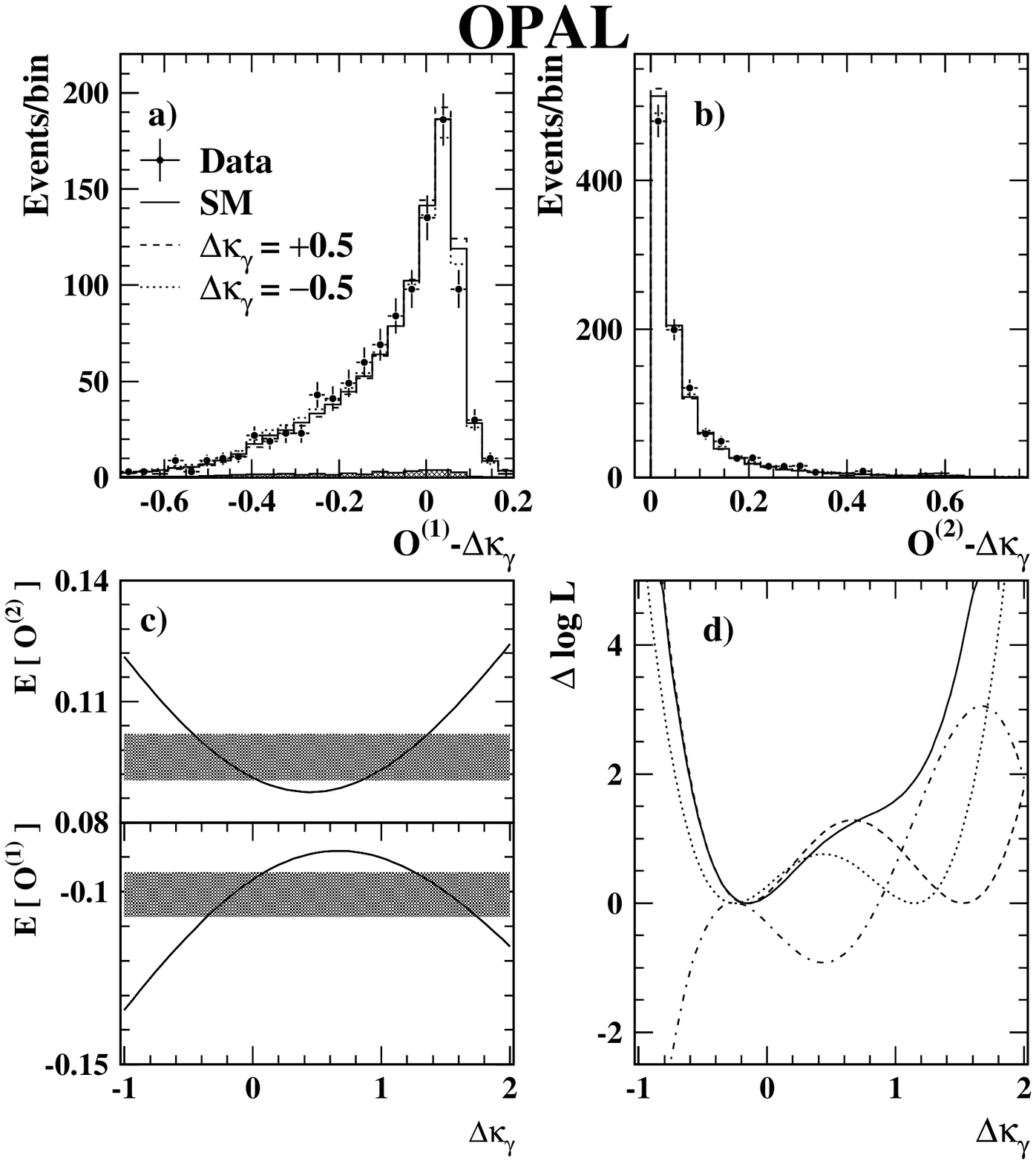,width=\textwidth}
\caption{\sl
\Qqln\ event distributions of the (a) first and (b) second order optimal 
observables corresponding to \dkg. The histograms show the expectation 
of the \SM\ and when $\dkg=\pm 0.5$. The expected mean values of these 
observables are plotted in (c) (lower and upper plot respectively) as 
solid curves. The measured mean values and their 
statistical errors are indicated by the shaded horizontal bands. 
In figure (d) the contributions of the first and second order optimal 
observable to the \LL\ function, disregarding the correlation between 
them, are plotted by the dashed and dotted lines respectively.
The dash-dotted line is the contribution of the 
correlation between the two observables and the solid line is the 
overall \LL\ function which is the sum of the three contributions.
}
\label{fig:oo}
\end{figure}

An illustrative example is shown in Figure~\ref{fig:oo} for  
$\OO_{\tiny \dkg}^{(1)}$ and $\OO_{\tiny \dkg}^{(2)}$ using the 
\Qqln\ final state. Figures~\ref{fig:oo}(a) and \ref{fig:oo}(b) show the
distributions of these observables in the data compared with the 
\MC\ expectations at the \SM\ and at \dkg=$\pm$0.5. The data agree 
with the \SM\ distributions. In Figure~\ref{fig:oo}(c) the 
expected mean values \OOEXP{\tiny \dkg}{(1)} 
and \OOEXP{\tiny \dkg}{(2)} as functions of \dkg\ are shown along with 
the measured mean values $\MOO^{(1)}_{\tiny \dkg}$ and
$\MOO^{(2)}_{\tiny \dkg}$.
 
A $\chi^2$ fit of the measured mean values \MOO\ to the corresponding 
expectations \OOEXP{}{}(\al) is performed to extract the couplings 
from the data. The covariance  matrix for the mean values is calculated 
from the \MC\ events. 
In the calculation of the $\chi^2$ for the \Qqln\ channel, the 
contribution of the \Qqtn\ final state is calculated separately,
in order to account for the worse resolution of \Qqtn\ events compared
with \Qqen\ and \Qqmn\ final states.
The \LL\ curve is obtained by subtracting 
from the $\chi^2$ function its minimum value and dividing by two.
For illustration, the \LL\ function corresponding to the one-dimensional
fit to obtain \dkg\ from the \Qqln\ data using 
$\OO_{\tiny \dkg}^{(1)}$ and $\OO_{\tiny \dkg}^{(2)}$ is shown in
Figure~\ref{fig:oo}(d). This \LL\ function can be decomposed into
three contributions coming from $\OO_{\tiny \dkg}^{(1)}$,
$\OO_{\tiny \dkg}^{(2)}$ and the correlation between them. 
The first two contributions are defined disregarding 
the correlation. Each of them has two minima at
equal height. Therefore, using only one of the two observables 
to calculate the \LL\ function results in an 
ambiguous two-fold solution, whereas using both observables leaves only 
one solution. Figure~\ref{fig:oo}(d) also demonstrates the importance 
of the correlation between the two observables.

The method is tested with a large number of Monte Carlo subsamples 
corresponding to the \SM\ and to various \tgc s. The size of each 
subsample corresponds to the collected luminosity. 
Summing up the \LL\ curves from the 
different subsamples corresponding to the same coupling gives 
results which are consistent with the coupling value used in 
the Monte Carlo generation.
%The error interval for the measurement of the coupling is then defined 
%in the usual way, as the region where the \LL\ function has a value of 
%no more than 0.5. To test the reliability of this error estimate, the 
%fraction of subsamples where the correct value is within the error
%interval is calculated.
Also the distribution of the fit results is centred around the expected
value, but there are some non Gaussian tails. Therefore, it is important
to test the reliability of the error interval which is estimated in the 
usual way, to be the region where the \LL\ function has a value of no more 
than 0.5. This is done by calculating the fraction of subsamples where
the correct value is within the error interval.
The error estimate is considered to be reliable
if the calculated fraction is consistent with 68\%, otherwise the 
\LL\ functions corresponding to the subsamples are scaled down by the 
necessary factor to obtain 68\% of the subsamples within the error
interval. The same scale factor is then applied also on the corresponding
\LL\ function of the data. The values obtained for this scale factor are
between 0.95 and 1. 
%This correcting procedure turned out to
%be necessary only for the \dkg\ and \dgz\ parameters derived from the \Qqln\ 
%and \Lnln\ samples respectively, requiring in both cases a scale factor of 
%0.85. 

\renewcommand{\arraystretch}{1.2}
\begin{table}[bhtp]
\begin{center}
\begin{tabular}{|l|c|c|c|} \hline 
                           & \dkg    & \dgz    &  \lg    \\ \hline 
\underline{\Qqln}          &         &         &         \\
Without systematics        & \dkgsta & \dgzsta & \lamsta \\
Expected stat. uncertainty & \dkgexp & \dgzexp & \lamexp \\
Including systematics      & \dkgchi & \dgzchi & \lamchi \\ 
Including 161-183 GeV data & \dkgiod & \dgziod & \lamiod \\
Including event rate       & \dkgbot & \dgzbot & \lambot \\ \hline
\underline{\Qqqq}          &         &         &         \\
Without systematics        & \qqksta & \qqgsta & \qqlsta \\
Expected stat. uncertainty & \qqkexp & \qqgexp & \qqlexp \\ 
Including systematics      & \qqktot & \qqgtot & \qqltot \\ 
Including 172-183 GeV data & \qqkiod & \qqgiod & \qqliod \\
Including event rate       & \qqkbot & \qqgbot & \qqlbot \\ \hline
\underline{\Lnln}          &         &         &         \\ 
Without systematics        & \dklsta & \dglsta & \lalsta \\
Expected stat. uncertainty & \dklexp & \dglexp & \lalexp \\
Including systematics      & \dklchi & \dglchi & \lalchi \\ 
Including 183 GeV data     & \dkliod & \dgliod & \laliod \\
Including event rate       & \dklbot & \dglbot & \lalbot \\ \hline
\underline{\rm All final states}
                           &         &         &         \\ 
Including systematics      & \sysdkg & \sysdgz & \syslam \\ 
Including previous data    & \ioddkg & \ioddgz & \iodlam \\ 
Overall combined results   &         &         &         \\ 
(including event rate)     & \resdkg & \resdgz & \reslam \\ 
95\% C.L. limits           & \dkgint & \dgzint & \lamint \\ 
3-D fit results            & \dkgthd & \dgzthd & \lamthd \\ \hline
\end{tabular}
\caption{Measured values of the \tgc\ parameters for each final state
and the results after combining all final states. For the individual 
final states we also list the results before including the systematic
errors and the expected statistical errors. We also list the results
after combining with our previous data and after combining  
with the event rate information from all \Com\ energies. 
For the overall combined data we list also the corresponding
95\% C.L. limits. In the last row we present the results of a 
3-dimensional fit to the overall combined data
where all three \tgc\ parameters are allowed to vary simultaneously.}
\label{tab:res}
\end{center}
\end{table}
\renewcommand{\arraystretch}{1.0}

Performing one-dimensional fits to the 189~\GeV\ data, we obtain the 
results quoted in Table~\ref{tab:res}. The expected statistical errors, 
which are the average fit errors of the \MC\ subsamples, are also quoted 
in Table~\ref{tab:res}, demonstrating the different sensitivities for
the various final states and \tgc\ parameters.  
%The statistical errors obtained from the fit are close to the expected 
%values listed in the table which are estimated from the average 
%statistical errors of the \MC\ subsample fit results.

As a cross-check to the mean optimal observable method, two other fit 
methods are applied to the \Qqln\ events. The first one is the 
binned maximum likelihood method which has been used to analyse
our previous data and is fully described in 
\cite{tgc161-analysis,tgc172-analysis}. For the present data we use a 
distribution in all five phase-space variables. This method yields 
sensitivities comparable with those of our main analysis.

The second cross-check fit for the \Qqln\ results uses
an unbinned maximum likelihood. The contribution of each data 
event to the likelihood is the sum over all \MC\ events of their weights,
which are Gaussian functions of the distance in phase-space to
that data event. In this way, most of the contribution comes from the 
\MC\ events nearest in phase space to the data event. The \tgc\ dependence 
of the likelihood is obtained by using \MC\ events at different \tgc\ 
values. This analysis has been applied so far only in a three-dimensional
phase-space in \Cthw, \Cthstl\ and \Phistl, and the obtained 
sensitivities are inferior to those of our main method.

For the \Qqqq\ events our fit method is cross-checked with a binned 
maximum likelihood fit to the shape of the \Cthw\ distribution as 
done in our previous analysis~\cite{tgc183-analysis}. Since it uses only 
one phase-space variable, this method yields worse sensitivities than
our main analysis.

The results of all these cross-check analyses are compatible with
those of our mean value optimal observable method.   

%%%%%%%%%%%%%%%%%%%%%%%%%%%%%%%%%%%%%%%%%%%%%%%%%%%%%%%%%%%%%%%%%%%%%%%%%

\section{Study of systematics}
\label{sec:sys}

The systematic uncertainties assigned to the event shape \tgc\ results are 
listed in Table~\ref{tab:sys}. The following sources are considered. 
\begin{itemize}
\item[a)]Our analysis relies on the predictions of the \Excalibur\ 
  \MC\ generator. The \Grace\ and \Koralw\ generators have a
  different calculation of the matrix elements and a different treatment
  of ISR. The associated systematic errors are assessed by applying our 
  analysis to event samples generated with these \MC\ programs.
\item[b)]The \MC\ modelling of the efficiency dependence on the
  event shape is checked by comparing data and \MC\ distributions
  of kinematic variables which might be affected by the selection cuts,
  such as the lepton and jet polar angle and energy, as well as the
  angle between the lepton in \Qqln\ events and the closest jet.
  An agreement is found between data and \MC\ within the statistical
  errors which are used to estimate the systematic uncertainty
  due to this source. For the \Lnln\ final state the event statistics are
  not sufficient to perform this check. The uncertainty for these 
  two charged lepton events is taken as twice the uncertainty of the
  \Qqln\ events with one charged lepton. This conservative estimate 
  is based on the assumption that the \Qqln\ uncertainty in the efficiency
  is entirely related to the lepton identification.   
\item[c)]Uncertainties in estimating the background
  are determined by varying both its shape and normalisation.
  The shape predicted by \Pythia\ of the \ZGqq\ background
  which is the main one for the \Qqln\ and \Qqqq\ final states is
  replaced by that predicted by \Herwig. The normalisation is varied 
  within the background uncertainty as determined in the \xse\
  analysis~\cite{sigww189}. The contributions from two-photon 
  processes are much smaller and a conservative estimate of their 
  uncertainties is obtained by removing them from the analysis.    
\item[d)]The uncertainty due to jet and lepton reconstruction
  is investigated by considering the effects of smearing and 
  shifting jet parameters in Monte Carlo events. The resolutions
  in jet energy, $\cos\theta$ and $\phi$ are varied by 10\%, 5\% and
  3\% respectively. 
  Jet and lepton energies are shifted by 0.5\% and jet
  $\cos\theta$ values are shifted by 0.0003. The sizes of these
  variations have been obtained from extensive studies of 
  back-to-back jets in \Zz\ events collected during calibration runs.
  The same reconstruction uncertainties are also assumed for the $\tau$
  jets in \Qqtn\ events. The momentum resolution of electrons and muons and 
  their charge misassignment probability are modified by varying the 
  resolution in $q/p_{\scriptscriptstyle T}$ by 10\%. Here $q$ and 
  $p_{\scriptscriptstyle T}$ are the lepton charge and transverse momentum
  with respect to the beam. 
% For electrons and muons in \Lnln\ events, the 
% energies are shifted by 0.3\% and the resolutions in energy, 
% $\cos\theta$ and $\phi$ are varied by 3\%, 10\% and 5\% respectively.
\item[e)]The uncertainty in the beam energy of 20 MeV and the effect
  of our \MC\ samples being produced at a centre-of-mass energy
  360 MeV higher than the data are investigated by applying our 
  analysis to \MC\ samples generated at different centre-of-mass energies
  then those used for reference samples.
  The effect of the W-mass uncertainty is assessed in a similar way
  using \MC\ samples generated with different W-masses.
\item[f)]Possible dependences on fragmentation models
  are studied by comparing the results of the analysis when applied to a 
  Monte Carlo sample generated with \Grace\ using either 
  \Jetset\ or \Herwig\ for the fragmentation phase. The main effect is on the
  \Qqqq\ final state, where \Herwig\ predicts a higher probability of correct
  jet pairing and correct W charge assignment than \Jetset.
  This effect is partially due to the average charged multiplicity
  for u,d,s,c quarks being lower in our tuned version of \Herwig\ than 
  in \Jetset. Since the data are in agreement with \Jetset,
  we weight the \Herwig\ Monte Carlo events so as to reproduce the same 
  mean charged  multiplicity as \Jetset. The effect of this weighting is 
  a reduction of the systematic error by about 25-30\%
  for all couplings.
\item[g)]Bose-Einstein correlations (BEC) in \Qqqq\ events
  might affect the measured W charge distribution. We investigate this 
  effect with a \MC\ program simulating BEC via re-shuffling of final
  state momenta~\cite{BEC}. 
\item[h)]The jet reconstruction and
  the measured W charge distribution might also be affected by colour 
  reconnection. This effect is investigated  with several
  Monte Carlo samples generated according to the 
  \Ariadne~\cite{ARIADNE} and Sj{\"{o}}strand-Khoze~\cite{SK} models.
   The largest effect is observed in the \Ariadne~2 and \Ariadne~3 
  models. However, these models have been disfavoured by studies of
  three jet events in LEP Z$^0$ data~\cite{Lep1_3jets}. For this reason
  the current \Ariadne\ implementation of colour reconnection is not used 
  to assign a systematic uncertainty. Of the remaining models considered,
  the Sj{\"{o}}strand-Khoze model I produces the largest bias. 
\end{itemize}
\begin{table}[htbp]
\begin{center}
\begin{tabular}{|ll||c|c|c||c|c|c||c|c|c|}\hline 
  \multicolumn{2}{|c||}{Source}& \multicolumn{3}{c||}{ \Qqln } &
  \multicolumn{3}{c||}{ \Qqqq } & \multicolumn{3}{c|}{ \Lnln } \\ \cline{3-11}
  \multicolumn{2}{|c||}{ } & \dkg & \dgz & \lg & \dkg & \dgz & \lg &
                                            \dkg & \dgz & \lg  \\ \hline 
 a) & MC generator        &\ddkgMC &\ddgzMC &\dlamMC & 0.05   & 0.01   & 0.01
                                            &\ddklMC &\ddglMC &\dlalMC \\
 b) & Efficiency          &\ddkgEF &\ddgzEF &\dlamEF & 0.03   & 0.01   & 0.02
                                            &\ddklEF &\ddglEF &\dlalEF \\
 c) & Background          &\ddkgBG &\ddgzBG &\dlamBG & 0.04   & 0.02   & 0.03
                                            &\ddklBG &\ddglBG &\dlalBG \\
 d) & Reconstruction      &\ddkgJT &\ddgzJT &\dlamJT & 0.01  & 0.01  & -
                                            &\ddklJT &\ddglJT &\dlalJT \\
 e) & \Ebeam\ and \Mw     &\ddkgEB &\ddgzEB &\dlamEB & 0.05   & 0.01   & 0.03
                                            &\ddklEB &\ddglEB &\dlalEB \\
 f) & Fragmentation       &\ddkgFR &\ddgzFR &\dlamFR & 0.17   & 0.06   & 0.08 
                                            &   -    &   -    &    -   \\
 g) & BEC                 &   -    &   -    &   -    & 0.04   & 0.02   & 0.03
                                            &   -    &   -    &    -   \\
 h) & Colour reconnection &   -    &   -    &   -    & 0.01   & 0.01   & 0.03
                                            &   -    &   -    &    -   \\\hline
    & Combined            &\ddkgsu &\ddgzsu &\dlamsu & 0.20   & 0.07   & 0.10
                                            &\ddklsu &\ddglsu &\dlalsu \\\hline
\end{tabular}
\end{center}
\caption{ Contributions to the systematic uncertainties in the 
  determination of the \tgc\ parameters from the 189~\GeV\ event shape
  data for different final states.}
\label{tab:sys}
\end{table}

The dominant contributions to the systematic errors in different final 
states come from different sources, and therefore we neglect any 
possible correlations between the systematic errors from different 
final states.

The systematic uncertainties from all sources are combined in quadrature. 
The values in Table~\ref{tab:sys} are not directly used in the analysis. 
Instead we use  
the systematic uncertainties in the expected mean optimal observables.
Correlations between the systematic errors of different 
observables are also taken into account and the full systematic covariance
matrix is added to the statistical covariance matrix. The combined
covariance matrix is used in the $\chi^2$ fits to obtain the \tgc s.
In this way, the \tgc\ errors obtained from the fit are already
the combined (statistical and systematic) errors.
The results of these fits are listed in Table~\ref{tab:res}.

%%%%%%%%%%%%%%%%%%%%%%%%%%%%%%%%%%%%%%%%%%%%%%%%%%%%%%%%%%%%%%%%%%%%%%%%%%

\section{Combined TGC results}
\label{sec:combtgc}

The \LL\ functions obtained from the \tgc\ event shape analyses
including the systematic errors are combined with the curves 
obtained from our previous analysis at lower centre-of-mass 
energies~\cite{tgc161-analysis,tgc172-analysis,tgc183-analysis}.
The systematic errors from the results of the different centre-of-mass
energies are assumed to be uncorrelated. This assumption does not
affect the final results, since the small contribution of the 
systematic uncertainties to the total errors of the previous data 
(see Tables 8, 10 and 12 in~\cite{tgc183-analysis}) becomes completely 
negligible after combining with the present results.

\begin{figure}[htbp]
\epsfig{file=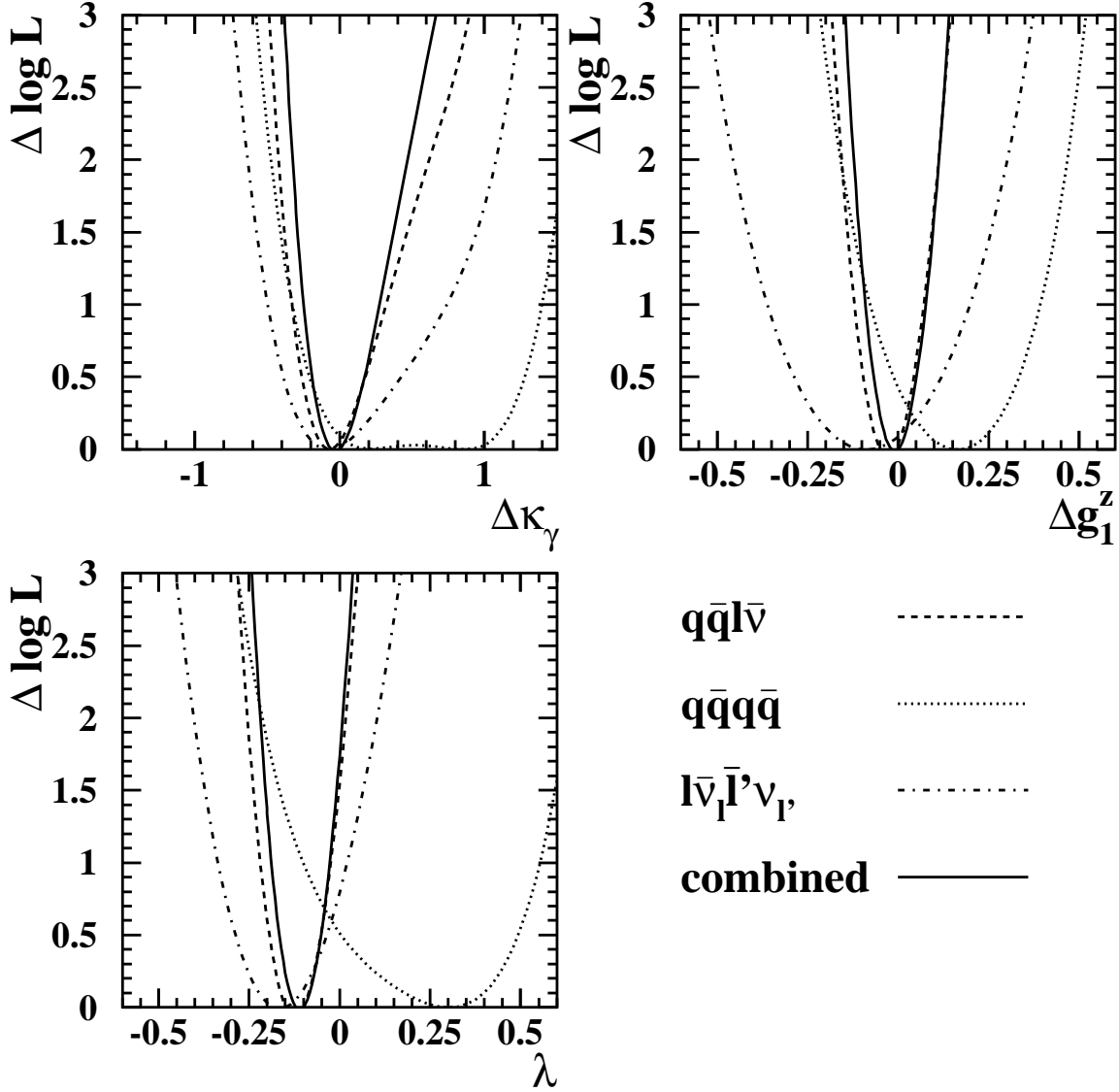,width=\textwidth}
\caption{\sl
Negative log-likelihood curves obtained from the different final  
states: \Qqln\ (dashed lines), \Qqqq\ (dotted lines) and \Lnln\
(dash-dotted lines). Each curve is obtained by combining the results
from the event shapes and the event rates and setting
the other two TGC parameters to their \SM\ values. Systematic errors 
are included. 
The solid line is obtained by combining the three final states.
}
\label{fig:channel}
\end{figure}

Table~\ref{tab:res} lists the event shape results for each final 
state after combining with the previous data. The corresponding
\LL\ functions are added to those obtained from the event rate.
The correlation between the systematic errors of the event rate and 
event shape, mainly due to the uncertainty in the background level,
is neglected, as it affects the results by less than 1\% of their 
statistical errors. The resulting \LL\ curves are shown in 
Figure~\ref{fig:channel} and the corresponding \tgc\ results are listed
in Table~\ref{tab:res}.
  
Finally the results of the three final states are combined.
This is done by first combining separately the three \LL\ functions 
from the event shapes and those from the event rate. In the next step
the event shape and event rate \LL\ functions are combined into 
the overall result. 
In combining the event shape information from the three
channels, the correlations between the systematic errors are neglected
as already explained in section~\ref{sec:sys}. On the other hand,
the correlations in the event rate data are not neglected,
as already mentioned in section~\ref{sec:rate}. 
The \LL\ curves obtained after combining the 
three final states are shown in Figure~\ref{fig:rateshape}, separately
for the event rate and event shape information and 
%in Figure~\ref{fig:channel} for 
their sum. In Table~\ref{tab:res} we list the \tgc\ results obtained 
from the combination over the three final states of the event shape
data and those after including the event rate information.
The overall combined results are shown also
in terms of 95\% C.L. intervals. 
%These are the combined results of this W-pair study. 
%Using our present and previous results we have \LL\ functions for all
%three final states. These are shown in Figure~\ref{fig:channel}.
%The \tgc\ results for the three event selections using the 
%angular distributions are combined by summing the corresponding 
%\LL\ functions.
%The correlations between the systematic errors of the three results 
%are neglected, since most of the important sources of systematic errors
%are relevant to a particular result and not common to all three of them.
%The \LL\ functions obtained for the different couplings are shown in
%Figure~\ref{fig:rateshape}. Adding these functions to those obtained
%using the total \xse\ yields the combined \LL\ functions
%which are plotted in Figure~\ref{fig:channel}.

%Finally our results can be further combined with those of our single W 
%analysis~\cite{single-analysis} based on the 161-183~\GeV\ data.
%The single-W process, \Wenu, has been investigated in the kinematic
%region where the outgoing electron goes undetected down the beam-pipe.
%For these events, the \xse\ is sensitive almost exclusively to 
%the WW$\gamma$ couplings~\cite{TSUKAMOTO}, namely \dkg\ and \lg. 
%From measuring the \xse, the \LL\ as a function of these two
%parameters has been obtained and is used here to combine with the 
%W pair results. The \LL\ functions following this combination are
%shown in Figure~\ref{fig:rateshape}, and the corresponding \tgc\ results
%and 95\% confidence level intervals are listed in Table~\ref{tab:combres}.

\begin{figure}[htbp]
\epsfig{file=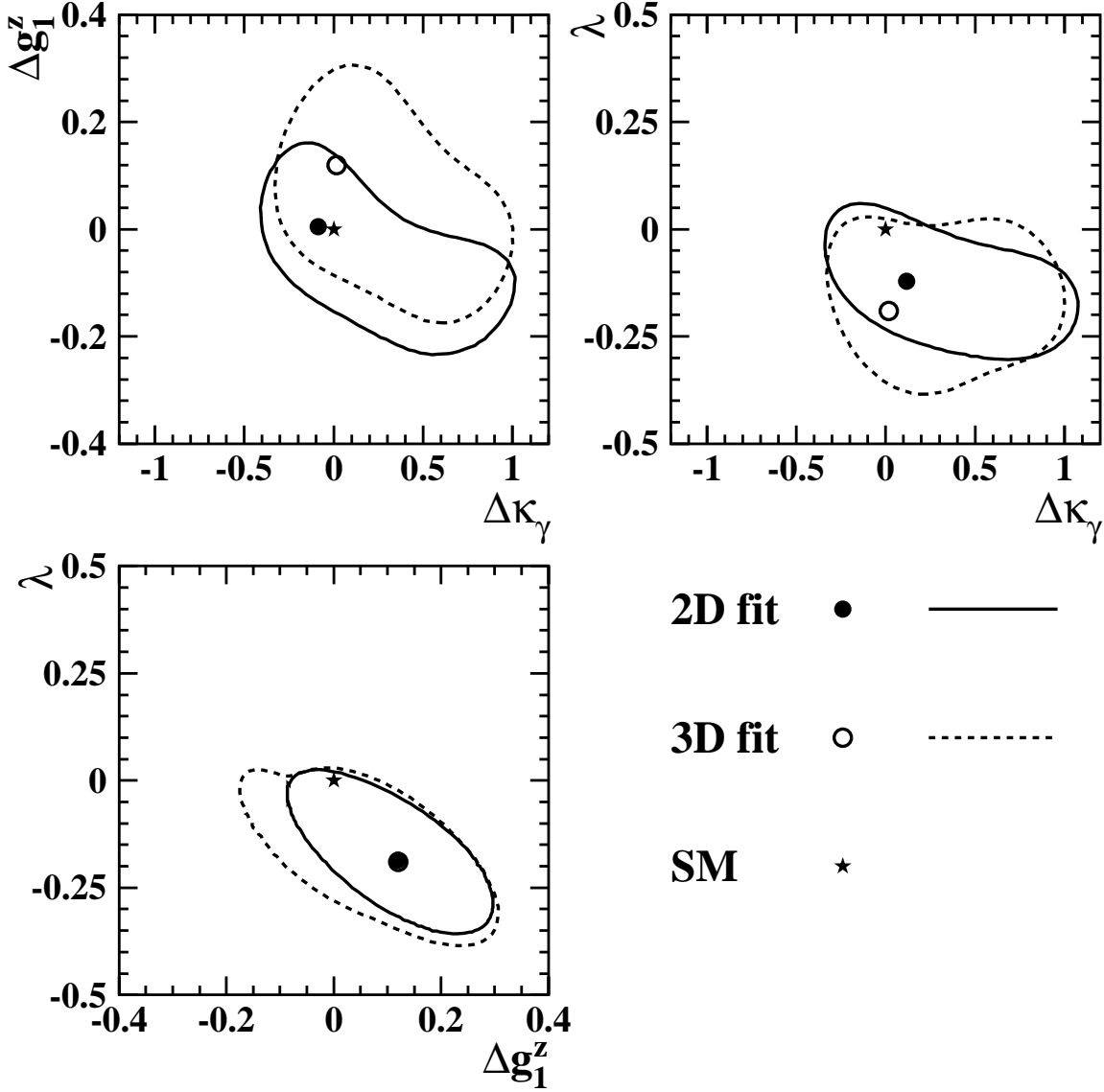,width=\textwidth}
\caption{\sl
The two-dimensional correlation contours corresponding to 
$\Delta\LL=2.995$ for different pairs of TGC parameters,
as obtained in two and three parameter fits.  The value 2.995 is 
chosen because in two dimensions it corresponds to a 95\% C.L. interval.
For the two parameter fit (solid curve), the third parameter is 
fixed at zero. In the case of the three parameter fit, 
the third parameter is not restricted and the dashed 
curve is a projection onto the two-dimensional plane of the envelope 
of the three dimensional $\Delta\LL=2.995$ surface.
The solid (open) points indicate the best 
two (three) parameter fit values and the stars indicate the \SM\ 
expectations. These results are obtained from all 
event rate data as well as event shapes of all \Qqln\ data, 
183--189 GeV \Qqqq\ data and 189 GeV \Lnln\ data. 
}
\label{fig:2d3d}
\end{figure}

To study correlations between the three \tgc\ parameters we also extract
a \LL\ as a function of all three variables, \dkg, \dgz\ and \lg.
Figure~\ref{fig:2d3d} shows the 95\% C.L. contour plots obtained from 
two-dimensional fits, where the third parameter is fixed at its
\SM\ value. We also perform a three-dimensional fit, where all three 
couplings are allowed to vary simultaneously. These fit results are listed
in the last row of Table~\ref{tab:res} and the corresponding two-dimensional
projections 
are plotted as dashed contour lines in Figure~\ref{fig:2d3d}. As can be 
seen, the allowed range for each parameter is extended when the constraints 
on the other two parameters are removed. The different minima locations
for the two- and three-dimensional fits in some of these plots, as well
as their non-elliptic shape, emerge from the presence of local minima 
in the three-dimensional \LL\ function~\cite{SEKULIN}.

%These three-dimensional fits can be used to obtain results for any other set
%of \tgc\ parameters, as long as relations~(\ref{su2u1}) between the five
%couplings are satisfied. For example, we determine the $\alpha$ parameters
%used in our previous publications~\cite{tgc161-analysis,tgc172-analysis}
%to be \abf=\resdkg, \awf=\resawf\ and \aw=\reslam. The result on each 
%$\alpha$ parameter is obtained assuming that the other two parameters 
%vanish.
%%%%%%%%%%%%%%%%%%%%%%%%%%%%%%%%%%%%%%%%%%%%%%%%%%%%%%%%%%%%%%%%%%%%%%%%%%

\section{Summary}
\label{sec:summary}  

Using a sample of \Ntotal\ \WW\ candidates collected at \LepII\ at a \Com\ 
energy of 189~\GeV, we measure the \tgc\ parameters. For this measurement 
we use the event rates and event shapes of W pairs decaying into all 
final states. After combining our results with those obtained from the 
161-183 GeV data we obtain:
\begin{eqnarray*}
\kg & = & \;\;\;\reskg, \\     
\gz & = & \;\;\;\resgz, \\    
\lg & = & \reslam,    
\end{eqnarray*}
where each parameter is determined when the other two parameters are set 
to their \SM\ values. These results supersede those from our previous 
publications~\cite{tgc161-analysis,tgc172-analysis,tgc183-analysis}.
They are all consistent with the \SM.

%%%%%%%%%%%%%%%%%%%%%%%%%%%%%%%%%%%%%%%%%%%%%%%%%%%%%%%%%%%%%%%%%%%%%%%%%%

\section*{Acknowledgements:}
%\par
%Acknowledgements:
%\par
We particularly wish to thank the SL Division for the efficient operation
of the LEP accelerator at all energies
 and for their continuing close cooperation with
our experimental group.  We thank our colleagues from CEA, DAPNIA/SPP,
CE-Saclay for their efforts over the years on the time-of-flight and trigger
systems which we continue to use.  In addition to the support staff at our own
institutions we are pleased to acknowledge the  \\
Department of Energy, USA, \\
National Science Foundation, USA, \\
Particle Physics and Astronomy Research Council, UK, \\
Natural Sciences and Engineering Research Council, Canada, \\
Israel Science Foundation, administered by the Israel
Academy of Science and Humanities, \\
Minerva Gesellschaft, \\
Benoziyo Center for High Energy Physics,\\
Japanese Ministry of Education, Science and Culture (the
Monbusho) and a grant under the Monbusho International
Science Research Program,\\
Japanese Society for the Promotion of Science (JSPS),\\
German Israeli Bi-national Science Foundation (GIF), \\
Bundesministerium f\"ur Bildung und Forschung, Germany, \\
National Research Council of Canada, \\
Research Corporation, USA,\\
Hungarian Foundation for Scientific Research, OTKA T-029328, 
T023793 and OTKA F-023259.\\

%%%%%%%%%%%%%%%%%%%%%%%%%%%%%%%%%%%%%%%%%%%%%%%%%%%%%%%%%%%%%%%%%%%%%%%%%%
 
\bibliography{pr324}

\begin{thebibliography}{10}

\bibitem{tgc161-analysis}
\opalacker, \PLB{397}{1997}{147}.

\bibitem{tgc172-analysis}
\opalacker, \EPC{2}{1998}{597}.

\bibitem{tgc183-analysis}
\opalabbien, \EPC{8}{1999}{191}.

\bibitem{OTHERLEPWW-tgc}
DELPHI Collaboration, P.\ Abreu \etal, \PLB{397}{1997}{158}; \\ L3
  Collaboration, M.\ Acciarri \etal, \PLB{398}{1997}{223}.; \\ L3
  Collaboration, M.\ Acciarri \etal, \PLB{413}{1997}{176}; \\ DELPHI
  Collaboration, P.\ Abreu \etal, \PLB{423}{1998}{194}; \\ ALEPH Collaboration,
  R.\ Barate \etal, \PLB{422}{1998}{369}; \\ DELPHI Collaboration, P.\ Abreu
  \etal, \PLB{459}{1999}{382}; \\ L3 Collaboration, M.\ Acciarri \etal,
  \PLB{467}{1999}{171}.

\bibitem{OTHERpr}
L3 Collaboration, M.\ Acciarri \etal, \PLB{403}{1997}{168}; \\ L3
  Collaboration, M.\ Acciarri \etal, \PLB{436}{1998}{417}; \\ ALEPH
  Collaboration, R.\ Barate \etal, \PLB{445}{1998}{239}; \\ ALEPH
  Collaboration, R.\ Barate \etal, \PLB{462}{1999}{389}; \\ L3 Collaboration,
  M.\ Acciarri \etal, \PLB{487}{2000}{229}.

\bibitem{Tevatron-tgc}
CDF Collaboration, F.\ Abe \etal, \PRL{78}{1997}{4536}; \\ D0 Collaboration,
  B.\ Abbott \etal, \PRD{60}{1999}{072002}.

\bibitem{LEP2YR}
Physics at LEP2, edited by G.\ Altarelli, T.\ Sj{\"{o}}strand and F.\ Zwirner,
  CERN 96-01 Vol. 1, 525.

\bibitem{HAGIWARA}
K.\ Hagiwara, R.D.\ Peccei, D.\ Zeppenfeld and K.\ Hikasa,
  \NPB{282}{1987}{253}.

\bibitem{BILENKY}
M.\ Bilenky, J.L.\ Kneur, F.M.\ Renard and D.\ Schildknecht,
  \NPB{409}{1993}{22}; \NPB{419}{1994}{240}.

\bibitem{GAEMERS}
K.\ Gaemers and G.\ Gounaris, \ZPC{1}{1979}{259}.

\bibitem{DERUJULA}
A.\ De Rujula, M.B.\ Gavela, P.\ Hernandez and E.\ Masso, \NPB{384}{1992}{3}.

\bibitem{HISZ}
K.\ Hagiwara, S.\ Ishihara, R.\ Szalapski and D.\ Zeppenfeld,
  \PLB{283}{1992}{353}; \PRD{48}{1993}{2182}.

\bibitem{SEKULIN}
R.L.\ Sekulin, \PLB{338}{1994}{369}.

\bibitem{OPAL}
OPAL Collaboration, K.\ Ahmet \etal, \NIMA{305}{1991}{275}; \\ M.\ Arignon
  \etal, \NIMA{313}{1992}103; \\ D.G\ Charlton, F.\ Meijers, T.J.\ Smith and
  P.S.\ Wells, \NIMA{325}{1993}129; \\ M.\ Baines \etal, \NIMA{325}{1993}271;
  \\ M.\ Arignon \etal, \NIMA{333}{1993}320; \\ B.E.\ Anderson \etal, IEEE
  Transactions on Nuclear Science, \textbf{41} (1994) 845; \\ S.\ Anderson
  \etal, \NIMA{403}{1998}{326}; \\ G.\ Aguillion \etal, \NIMA{417}{1998}{266}.

\bibitem{GOPAL}
J.\ Allison \etal, \NIMA{317}{1992}{47}.

\bibitem{EXCALIBUR}
F.A.\ Berends, R.Pittau and R.\ Kleiss, \CPC{85}{1995}{437};\\ F.A.\ Berends
  and A.I.\ van Sighem, \NPB{454}{1995}{467}.

\bibitem{GRC4F}
J.\ Fujimoto \etal, \CPC{100}{1997}{128}.

\bibitem{KORALW}
M.\ Skrzypek \etal, \CPC{94}{1996}{216}; \\ M.\ Skrzypek \etal,
  \PLB{372}{1996}{289}; \\ S.\ Jadach \etal, \CPC{119}{1999}{272}.

\bibitem{PYTHIA}
T.\ Sj{\"{o}}strand, \CPC{82}{1994}{74}.

\bibitem{HERWIG}
G.\ Marchesini \etal, \CPC{67}{1992}{465}.

\bibitem{KORALZ}
S.\ Jadach \etal, \CPC{79}{1994}{503}.

\bibitem{BHWIDE}
S.\ Jadach, W.\ Placzek and B.F.L.\ Ward, \PLB{390}{1997}{298}.

\bibitem{PHOJET}
R.\ Engel, \ZPC{66}{1995}{203}; \\ R.\ Engel and J.\ Ranft,
  \PRD{54}{1996}{4244}.

\bibitem{VERMASEREN}
J.A.M.\ Vermaseren, \NPB{229}{1983}{347}.

\bibitem{sigww189}
\opalabbien, ``W$^+$W$^-$ production cross-section and W branching fractions in
  \epem\ collisions at 189 GeV, CERN-EP/2000-101'', submitted to Phys. Lett. B.

\bibitem{MWPDG}
D0 Collaboration, B.\ Abbott \etal, \PRL{84}{2000}{222}; \\ CDF Collaboration,
  F.\ Abe \etal, \PRL{75}{1995}{11}.

\bibitem{RacoonWW}
A.\ Denner \etal, \PLB{475}{2000}{127}; \\ A.\ Denner \etal, ``Electroweak
  radiative corrections to $\epem\rightarrow{\rm WW}\rightarrow$ four fermions
  in double pole approximation -- the \Racoon\ approach'', BI-TP 2000/06, {\tt
  http://arXiv.org/abs/hep-ph/0006307}.

\bibitem{YFSWW}
S.\ Jadach \etal, \PRD{61}{2000}{113010}; \\ S.\ Jadach \etal, ``Precision
  predictions for (un)stable \WW\ pair production at and beyond LEP2
  energies'', {\tt http://arXiv.org/abs/hep-ph/0007012}, UTHEP-00-0101, Jan.
  2000, submitted to Phys. Lett. B.

\bibitem{FFMCWG}
M.\ Gr{\"{u}}newald \etal, ``Four-fermion production in electron-positron
  collisions'', The LEP-2 Monte Carlo Workshop 1999/2000, {\tt
  http://arXiv.org/abs/hep-ph/0005309}.

\bibitem{GENTLE}
D.\ Bardin \etal, Nucl. Phys. B, Proc. Suppl. {\bf 37B} (1994) 148; \\ D.\
  Bardin \etal, \CPC{104}{1997}{161}.

\bibitem{DURHAM}
N.\ Brown and W.J.\ Stirling, \PLB{252}{1990}{657}; \\ S.\ Catani \etal,
  \PLB{269}{1991}{432}; \\ S.\ Bethke, Z.\ Kunszt, D.\ Soper and W.J.\
  Stirling, \NPB{370}{1992}{310}; \\ N.\ Brown and W.J.\ Stirling,
  \ZPC{53}{1992}{629}.

\bibitem{GCE}
\opalakrawy, \PLB{253}{1991}{511}.

\bibitem{MT}
\opalacker, \EPC{2}{1998}{213}.

\bibitem{Karlen}
D.\ Karlen, Computers in Physics {\bf 12} (1998) 380, \\ {\tt
  http://arXiv.org/abs/physics/9805018}.

\bibitem{Fanourakis}
G.K.\ Fanourakis, D.\ Fassouliotis and S.E.\ Tzamarias, \NIMA{412}{1998}{465};
  \NIMA{414}{1998}{399}.

\bibitem{Rouge}
M.\ Davier, L.\ Duflot, F.\ LeDiberder and A.\ Roug\'{e}, \PLB{306}{1993}{411}.

\bibitem{Diehl}
M.\ Diehl and O.\ Nachtmann, \ZPC{62}{1994}{397}.

\bibitem{Fanourakis2}
G.K.\ Fanourakis \etal, \NIMA{430}{1999}{455}; \NIMA{430}{1999}{474}.

\bibitem{BEC}
L.\ L{\"{o}}nnblad and T.\ Sj{\"{o}}strand, \EPC{2}{1998}{165}.

\bibitem{ARIADNE}
L.\ L{\"{o}}nnblad, \CPC{71}{1992}{15}.

\bibitem{SK}
T.\ Sj{\"{o}}strand and. V.A.\ Khoze, \ZPC{62}{1994}{281}; \PRL{72}{1994}{28}.

\bibitem{Lep1_3jets}
\opalabbien, \EPC{11}{1999}{217}.

\end{thebibliography}

%%%%%%%%%%%%%%%%%%%%%%%%%%%%%%%%%%%%%%%%%%%%%%%%%%%%%%%%%%%%%%%%%%%%%%%%%

%%%%%%%%%%%%%%%%%%%%%%%%%%%%%%%%%%%%%%%%%%%%%%%%%%%%%%%%%%%%%%%%%%%%%%%%

%%%%%%%%%%%%%%%%%%%%%%%%%%%%%%%%%%%%%%%%%%%%%%%%%%%%%%%%%%%%%%%%%%%%%%%%

%%%%%%%%%%%%%%%%%%%%%%%%%%%%%%%%%%%%%%%%%%%%%%%%%%%%%%%%%%%%%%%%%%%%%%%%%

%%%%%%%%%%%%%%%%%%%%%%%%%%%%%%%%%%%%%%%%%%%%%%%%%%%%%%%%%%%%%%%%%%%%%%%%%

%%%%%%%%%%%%%%%%%%%%%%%%%%%%%%%%%%%%%%%%%%%%%%%%%%%%%%%%%%%%%%%%%%%%%%%%

%%%%%%%%%%%%%%%%%%%%%%%%%%%%%%%%%%%%%%%%%%%%%%%%%%%%%%%%%%%%%%%%%%%%%%%%%%%

%%%%%%%%%%%%%%%%%%%%%%%%%%%%%%%%%%%%%%%%%%%%%%%%%%%%%%%%%%%%%%%%%%%%%%%%%

%%%%%%%%%%%%%%%%%%%%%%%%%%%%%%%%%%%%%%%%%%%%%%%%%%%%%%%%%%%%%%%%%%%%%%%
\end{document}